\RequirePackage{lineno}
\documentclass[%
 onecolumn,
 amsmath,amssymb,
 aps,
]{revtex4-2}
\usepackage[margin=1in]{geometry}
\usepackage{graphicx}
\usepackage{dcolumn}
\usepackage{bm}
\usepackage{subcaption}

\usepackage{amsmath, amssymb}
\usepackage{setspace}
\newcommand{\bx}{{\mathbf x}}
\newcommand{\bxs}{{\mathbf{x}^*}}
\newcommand{\bbD}{{\mathbb D}}
\newcommand{\bbR}{{\mathbb R}}
\newcommand{\bC}{{\mathbf C}}

\newcommand{\bc}{{\mathbf c}}

\newcommand{\bX}{{\mathbf X}}
\usepackage{color}

\begin{document}
\preprint{}

\title{Physics-informed CoKriging model of a redox flow battery}

\author{Amanda A. Howard}
\affiliation{Pacific Northwest National Laboratory, Richland, WA}

\author{Alexandre M. Tartakovsky}
\affiliation{Pacific Northwest National Laboratory, Richland, WA; Department of Civil and Environmental Engineering, University of Illinois Urbana-Champaign, Urbana, IL}
\email{amt1998@illinois.edu}

\date{\today}

\begin{abstract}

Redox flow batteries (RFBs) offer the capability to store large amounts of energy cheaply and efficiently, however, there is a need for fast and accurate models of the charge-discharge curve of a RFB to potentially improve the battery capacity and performance. We develop a multifidelity model for predicting the charge-discharge curve of a RFB. In the multifidelity model, we use the Physics-informed CoKriging (CoPhIK) machine learning method that is trained on experimental data and constrained by the so-called ``zero-dimensional'' physics-based model. Here we demonstrate that the model shows good agreement with experimental results and significant improvements over existing zero-dimensional models. We show that the proposed model is robust as it is not sensitive to the input parameters in the zero-dimensional model. We also show that only a small amount of high-fidelity experimental datasets are needed for accurate predictions for the range of considered input parameters, which include current density, flow rate, and initial concentrations. 
\end{abstract}

\maketitle

In this paper, we develop a multifidelity model for a redox flow battery (RFB) that incorporates both experimental and physics-based modeling results. RFBs are among the most promising flow batteries, and accurate modeling is necessary to predict and further improve the performance of such systems. 

Recent advances in machine learning have begun to make way for accurate and reliable predictions of the performance of batteries. Much of the recent work has focused on lithium-ion batteries, including predicting the long-term performance of such batteries over hundreds of cycles \cite{liu2015, severson2019, weigert2011} and using machine learning for material development \cite{sanchez2018, wu2018}. Work on RFBs is more limited. \cite{li2020} used machine learning approach to optimize the cost and performance of a vanadium flow battery, and \cite{bao2020} uses a deep neural network to couple pore-scale microstructure simulations and device-scale performance. In all cases, a key limitation to machine learning approaches for battery design is the availability of large, robust datasets for training \cite{ severson2019, li2020}. Experimental datasets are expensive and time-consuming to produce \cite{li2020}, while datasets generated from simulations are too computationally expensive to represent the full device at high accuracy \cite{bao2020}. In this work, we propose using a computationally efficient physics-based model to reduce the amount of experimental data needed for training. 

Several physics-based models of RFB systems have been proposed in recent years, including the zero-dimensional (0d) lumped-parameter model \cite{Shah2011}. This control-oriented model is derived for a unit cell RFB system and contains several fitting parameters. These parameters are determined by fitting the 0d model to experiments and vary widely between experiments even for the same battery design \cite{Shah2011, Eapen2019}. 
The dependence of parameters on initial and operating conditions reduces the predictive ability of the 0d model. 

In this work, we extend the multifidelity Physics-Informed CoKriging method, which was originally introduced in \cite{yang2019} and termed as  CoPhIK. This approach is based on the Physics-Informed Kriging (PhIK) \cite{tartakovsky2019physicsHICSS,yang2018}, where unknown coefficients in the physics model, in this case the fitting parameters for the 0d model, are replaced with Gaussian random variables. This turns the 0d model in the stochastic model used to compute covariances for the Gaussian process regression (GPR) \cite{rasmussen2003gaussian} model of the RFB. PhIK provides predictions of both the most probable  behavior of the system (in the form of the posterior mean) and the uncertainty in the prediction (in the form of posterior variance).  
However, PhIK is limited to the accuracy of the underlying physics model. CoPhIK overcomes this limitation by using a small experimental dataset to establish the correlation between the physics-based model and the behavior of a real system to improve the model's predictions. We are able to accurately capture the performance of cell voltage over a charge-discharge cycle. CoPhIK is computationally cheap to run, with a typical prediction requiring less than a minute to run and requires no computationally expensive pre-training. Additionally, CoPhIK produces accurate results with a very small experimental dataset, requiring only fifteen experiments are used for training.

\section{Experimental data}\label{data}

In this paper, we use experimental data from RFB laboratory experiments as ``high-fidelity'' data. We also use data generated by the 0d model that we refer to as ``low-fidelity'' data because of the strong assumptions and simplifications in the 0d model. In this section, we detail the high-fidelity dataset. The low-fidelity data set and 0d model are described in the Methods section. 
\begin{figure}[h]
\centering
\includegraphics[width=0.6\linewidth]{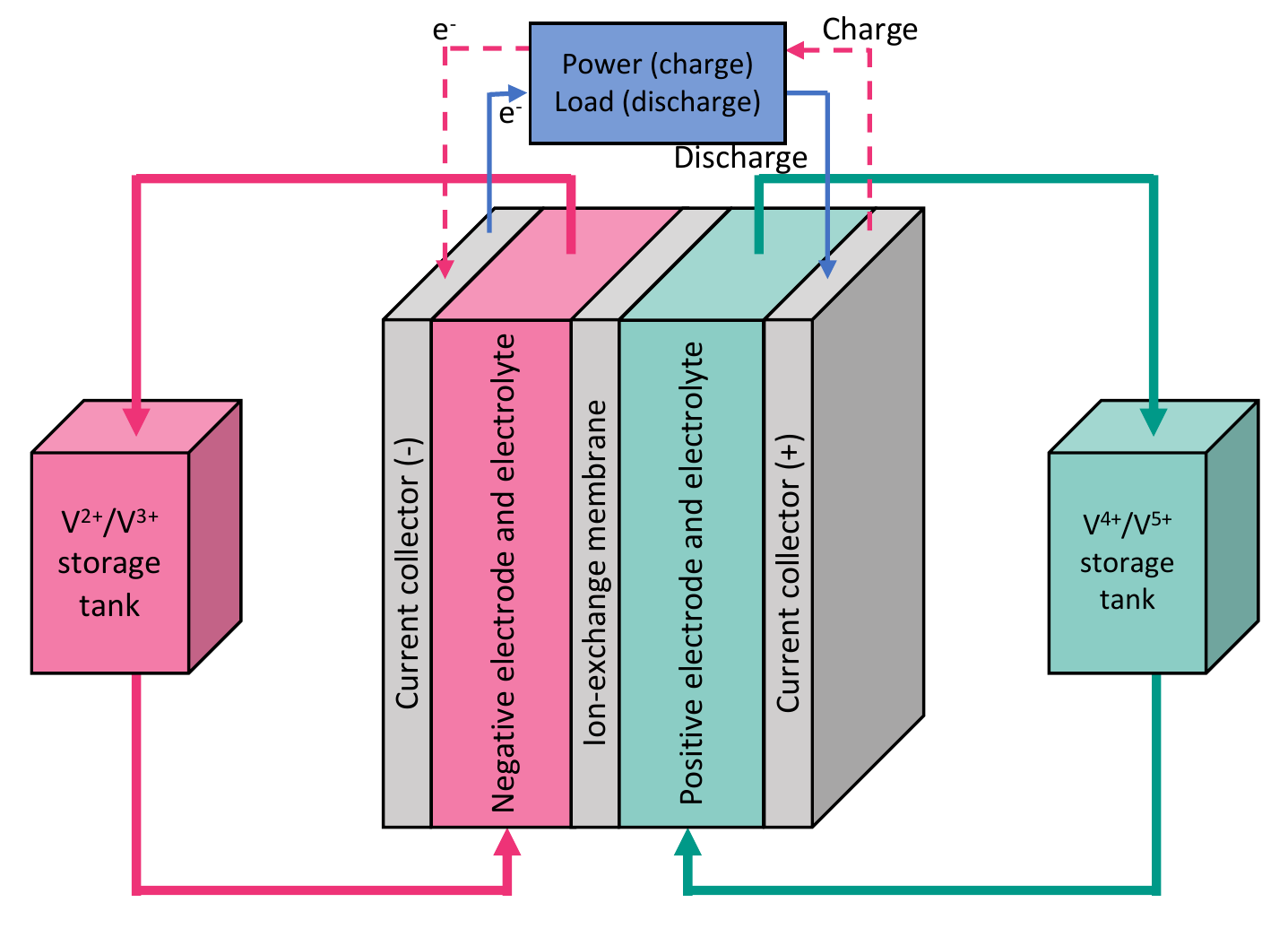}
\caption{Schematic of the unit cell battery used in the experiments from \cite{bao2020}. }\label{fig:battery}
\end{figure}

The experimental dataset is collected in sixteen RFB experiments under conditions listed in the Supplementary Material.
 A schematic of the unit cells used in the experiments is shown in Figure \ref{fig:battery}. 
  While many cycles were completed for each experiment, here we consider the performance only of the third cycle for training and validation as we did not observe any significant changes in the battery performance in the following cycles. The first two cycles are discarded as ``priming'' cycles during which batteries do not reach the expected capacity.  

\begin{figure}[h]
\centering
\begin{subfigure}{0.24\textwidth}
\includegraphics[width=\textwidth]{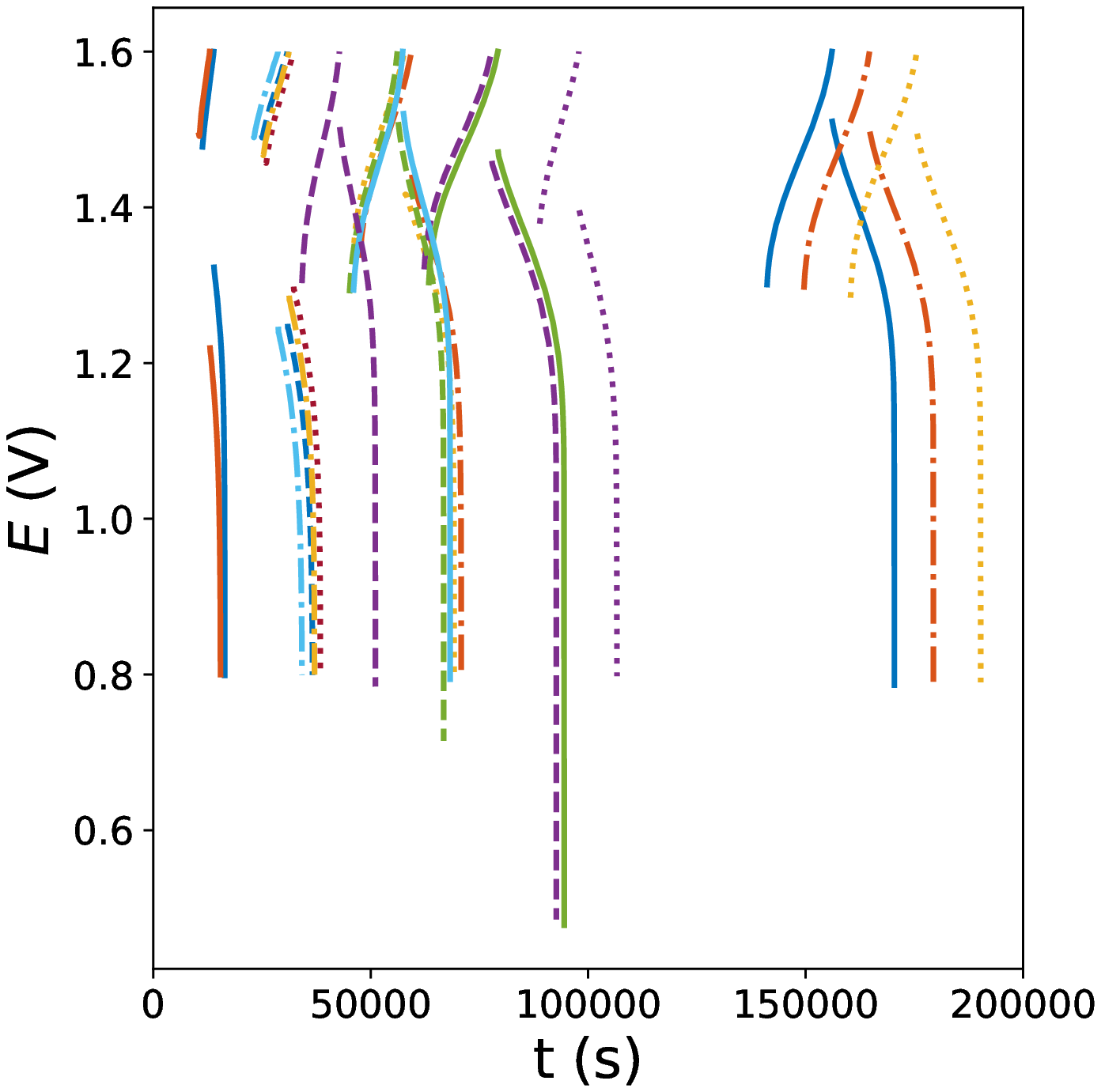}
\caption{}
\end{subfigure}
\begin{subfigure}{0.24\textwidth}
\includegraphics[width=\textwidth]{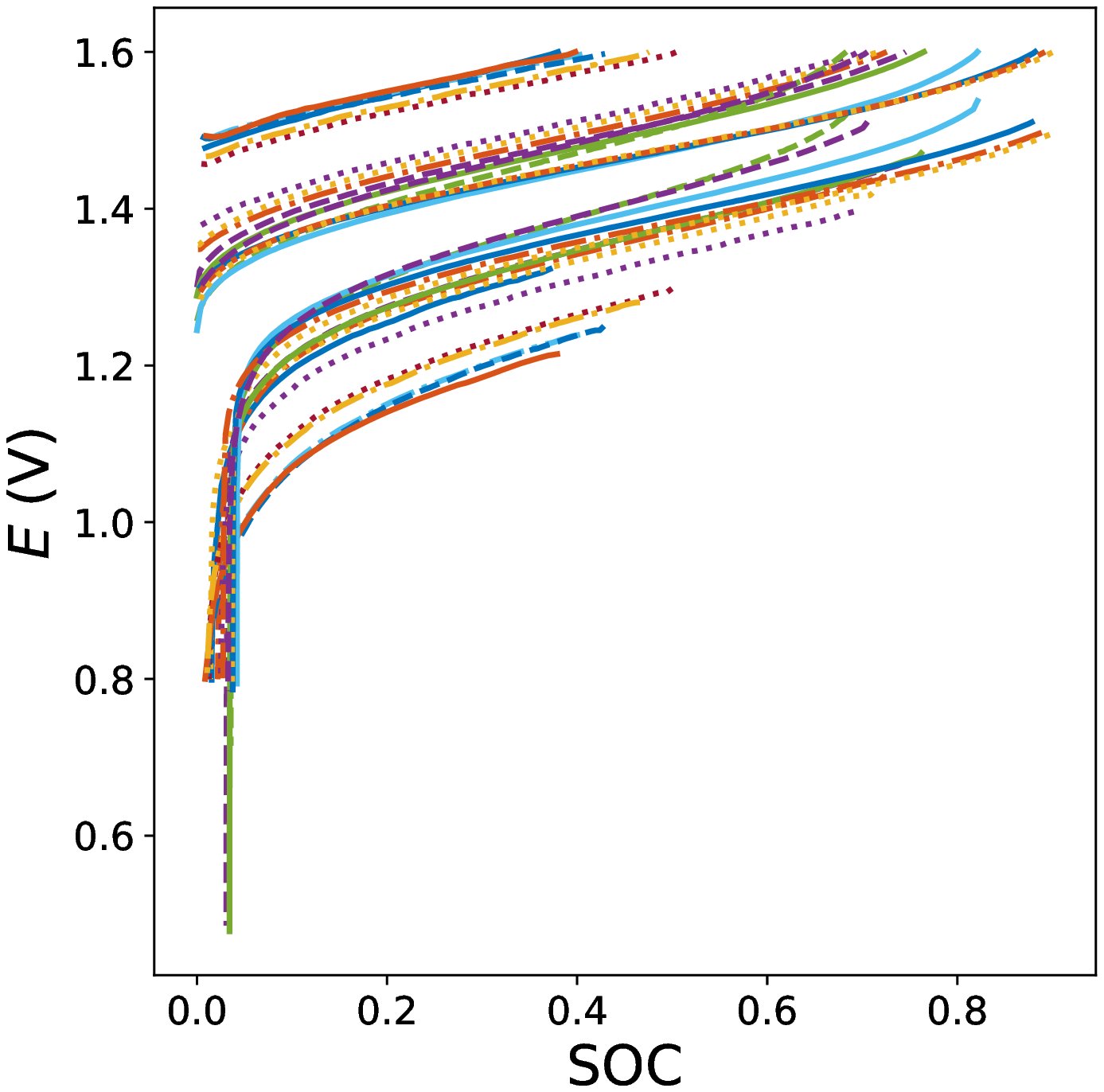}
\caption{}
\end{subfigure}
\begin{subfigure}{0.24\textwidth}
\includegraphics[width=\textwidth]{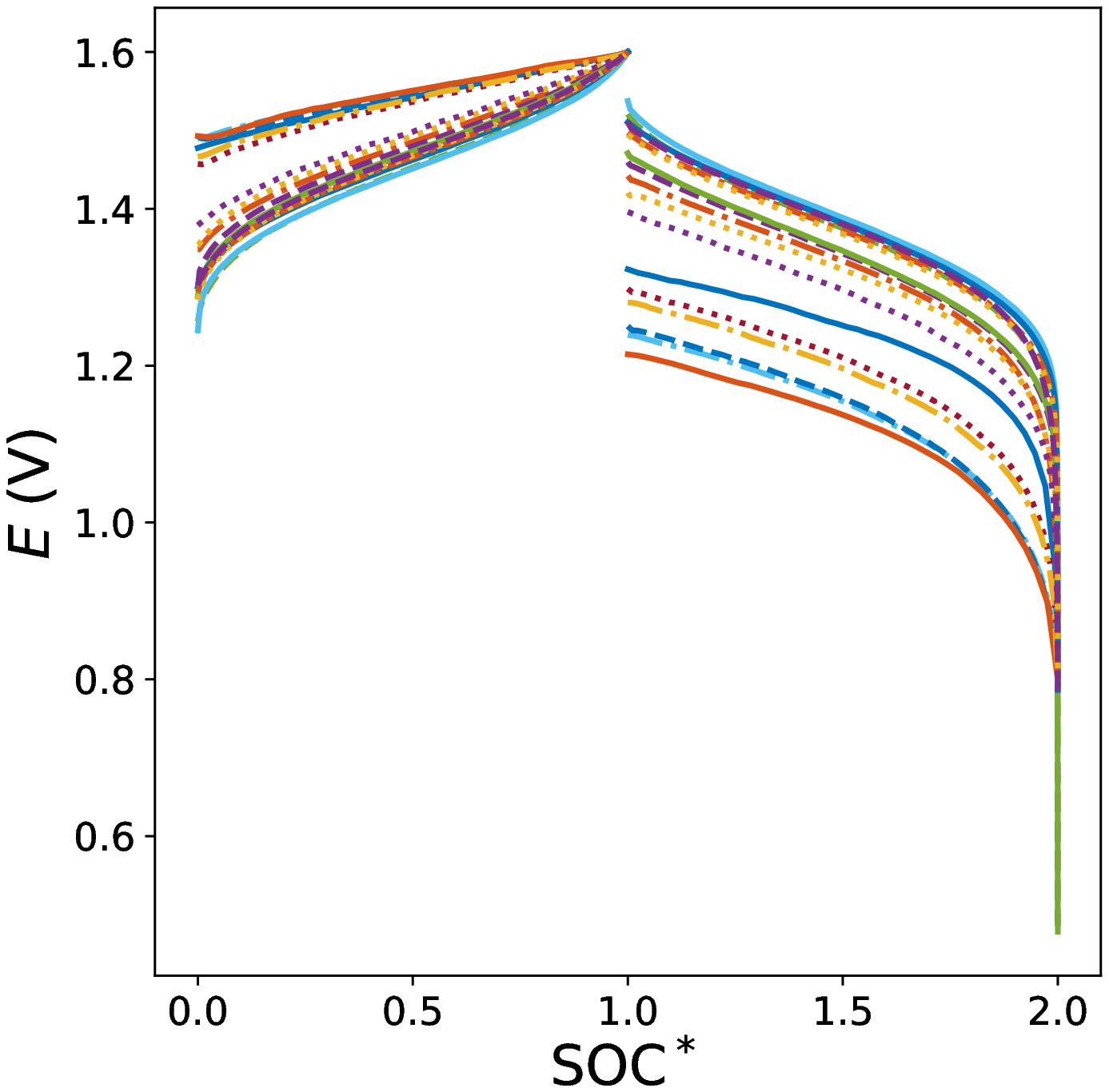}
\caption{}
\end{subfigure}
\begin{subfigure}{0.24\textwidth}
\includegraphics[width=\textwidth]{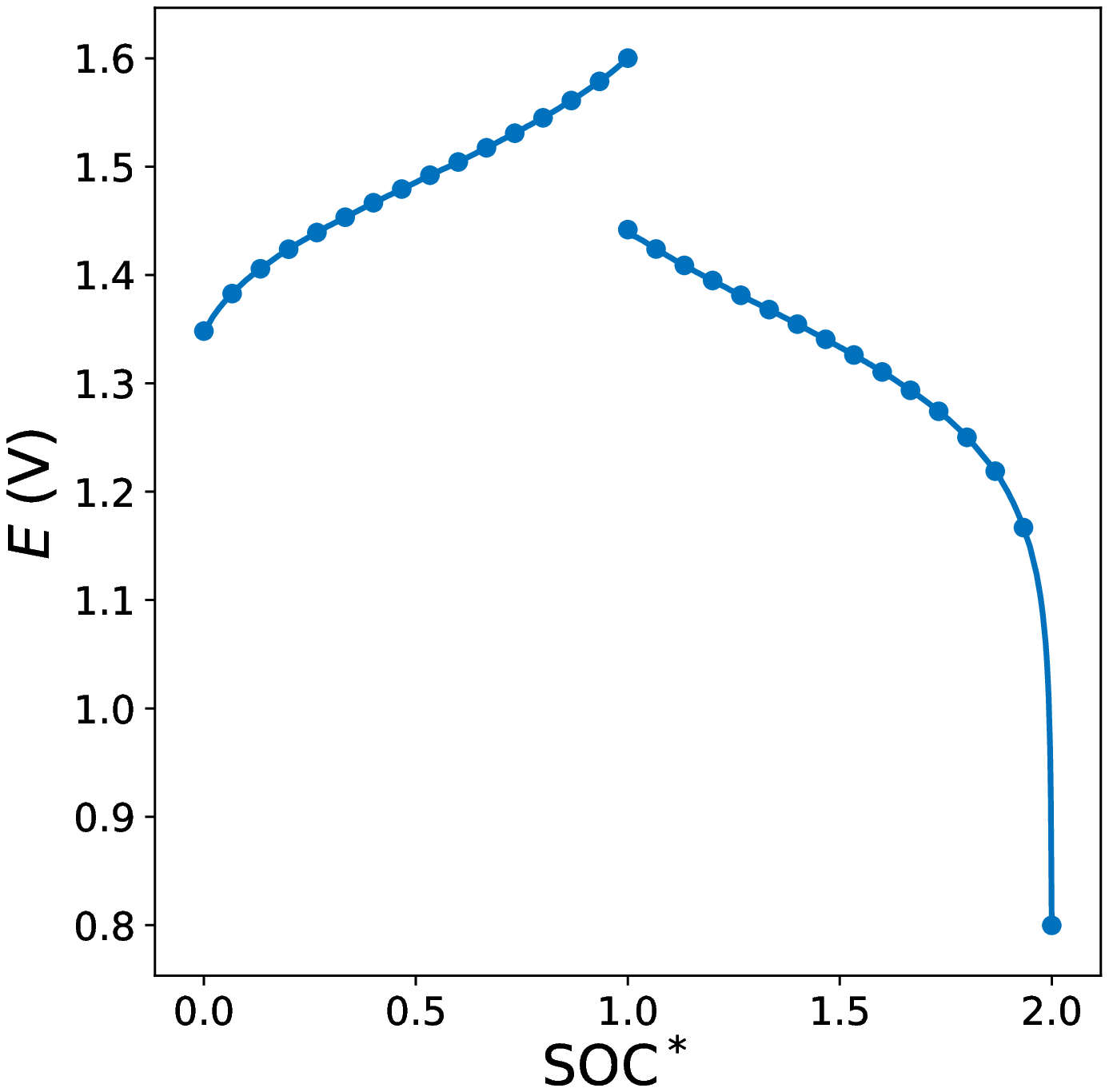}
\caption{}
\end{subfigure}
\caption{(a) All experimental data plotted with time $t$. (b) All experimental data with SOC. (c) All experimental data with scaled SOC, $SOC^*$. (d) One example of the interpolated fit. For clarity, only a subset of points in the interpolation vector $\mathbf{s^*}$ are shown.} \label{fig:hf_data}
\end{figure}

In all experiments the cell potential, $E$, was measured as a function of time during charging and discharging cycles as shown in Fig. \ref{fig:hf_data}a. In our model, it is convenient to express concentration as a function of the state of charge (SOC), see Fig. \ref{fig:hf_data}b. The SOC varies between zero for a fully discharged battery and one for a fully charged battery and can be found as 
$ SOC(t) = \frac{C_{V(II)}(t)}{C_{V_n}^0} $
where $C_i(t)$ is the concentration of species $i$ at time $t$ and $C_i^0$ is the initial concentration of species $i$. $C_{V_n}^0$ is the total initial vanadium concentration of the initial half-cell and $C_{V(II)}(t)$ is the concentration of species V(II) at time $t$  \cite{Chen2014a}.  
To validate our approach, we model each experiment using data from the remaining fifteen experiments as a training set for the CoPhIK model (described in Section \ref{cophik}).
One challenge with the considered experimental dataset is that the experiments occur on very disparate timescales resulting in the different values of SOC at the end of the cycles, see Figs. \ref{fig:hf_data}a and b. 
Therefore, we rescale the SOC for each experiment so that the charging cycle has range $[0, 1]$ and the discharge cycle has range $[1, 2]$ through the following operations: 

\begin{align}
    SOC^*_C &= (SOC_C-SOC_C(0))/(\max(SOC_C)-SOC_C(0)) \label{eq:SOC_scale1} \\
    SOC^*_D &= 1+ (SOC_D-SOC_D(0))/(\max(SOC_D)-SOC_D(0)) \label{eq:SOC_scale2}
\end{align}
where the subscripts $C$ and $D$ denote charging and discharging, respectively. 
With this change, the experiments collapse onto a narrow manifold, see Fig. \ref{fig:hf_data}c.

Finally, for the CoPhIK model  we need data at the same (rescaled) SOC instances for all the experiments. To obtain such data, we select an SOC vector $\mathbf{SOC^*} = [0, s_1^*, s_2^*, ..., 2.0]$ of length $N_s$ and interpolate the experimental results onto the $s_i^*$ components, $i=1,2,...N_s$, of the vector $\mathbf{s^*}$  with B-spline interpolation optimization performed using the  \texttt{interpolate} package in \texttt{scipy}, see Fig. \ref{fig:hf_data}d for an example of the interpolated fit for one of the experiments.

\section{Multifidelity framework}
\label{sec:Method}

\subsection{Gaussian process regression}

We begin by describing the GPR method for modeling cell potential $E$ of the RFB 
as a function of SOC and parameters describing the battery operation conditions and properties of the battery.
Defining the input space of the GPR model is an important step. The number of parameters describing the battery and its operating condition is very large, see the Supplementary Material.
Ideally, the input space of the GPR model would include all these parameters and SOC such that the GPR model can predict the cell voltage $E$ at any SOC for any type and size of the RFB. However, the training of such GPR model would require vast amounts of computational time and data, with both training time of the GPR model and required data exponentially increasing with the number of input parameters. Since we only have 16 experiments (15 experiments for training to account for a test data) we must reduce the number of input parameters as much as possible.  Therefore, we only include the varying operation conditions in the considered experiments and SOC in the input space and use the 0d model to describe the dependence of $E$ on the rest of parameters. Since all 16 experiments are performed with fixed velocity $v = 4.17$ m/s and concentrations $C_{V(II)}^0 = 0$ mol/m$^3$,  and $C_{V(V)}^0 = 0$ mol/m$^3$, we exclude $v$,  $C_{V(II)}^0$, and $C_{V(V)}^0$ from the input space. The resulting input space of the GPR model is 

$$\mathbb{D} = [s, j, V_r, w_m, C_{V(III)}^0, C_{V(IV)}^0, C_{H^+, pos}^0, C_{H_2O, pos}^0, C_{H^+, neg}^0, C_{H_2O, neg}^0] \subseteq \mathbb{R}^d,$$ with $d = 10$. The cell voltage, $E$, is the output (or state of interest) of the GPR model, $E \; : \; \mathbb{D} \rightarrow \mathbb{R}$. In other words, the GPR model aims to predict $E$ for any $s$, $j$, $V_r$, $w_m$, $C_{V(III)}^0$, $C_{V(IV)}^0$, $C_{H^+, pos}^0$, $C_{H_2O, pos}^0$, $C_{H^+, neg}^0$, and $C_{H_2O, neg}^0$ with fixed $v = 4.17$ m/s and $C_{V(II)}^0= C_{V(V)}^0 = 0$ mol/m$^3$. 

Next, we organize the measurements of $E$ at locations 
 $\mathbf{X} =  \{ {\mathbf{x}^{(i)}} \}_{i=1}^{N}$ 
in the observation vector  $\mathbf{E} = (E^{1}, E^{2}, \ldots, E^{N})^T$, where $N=15\cdot N_s$ is the total number of observations for fifteen experiments, $N_s$ is the size of the $\mathbf{s}^*$ vector of SOC measurements, and $E^{i}$ is the $E$ measurement at the location $\mathbf{x}^{(i)}$. 

In the GPR approach we treat $E(\mathbf{x})$ as a Gaussian process as described in the Methods section. Estimating the mean $\mu(\bx)$ and covariance $k(\bx, \bx')$ models is the main challenge in GPR-based methods (see Methods.) In the standard data-driven GPR,  $\mu(\bx)$ and $k(\bx, \bx')$ would be computed from data by minimizing the so-called marginal log-likelihood function ~\cite{williams2006gaussian}. It should be noted that the prior statistics (i.e., $\mu(\bx)$ and $k(\bx, \bx')$) computed in this approach do not satisfy the physical laws governing the system of interest unless the phase space is very densely sampled, which can lead to significant errors and nonphysical behavior \cite{tartakovsky2019physicsHICSS,yang2018}. To address this limitation of the data-driven GPR method, in the following Section \ref{phik} we propose to estimate the prior statistics from the 0d model where we treat unknown parameters as random variables. A similar approach was used in \cite{tartakovsky2019physicsHICSS,yang2018} in GPR models of subsurface transport and power grid dynamics.

\subsection{Estimating prior statistics from the stochastic 0d model}\label{phik}

The motivation behind this approach is that the 0d model, which is described in Methods, has the unknown parameters  $k_{1}, k_2, \sigma_e$, and $S$ denoting the reference rate constants, ionic concentration of the electrolyte, and specific surface area, which depend on operating conditions and must be calibrated for each experiment. In this work we treat these parameters as Gaussian random variables with prescribed (prior) mean and variance that turns the 0d model in a stochastic model. Next, we use the Monte Carlo (MC) method to compute the prior GP model of $E$ based on the stochastic 0d model. Specifically, we generate $N_{MC}$ realizations of the $k_{1}, k_2, \sigma_e$, and $S$ parameters and compute $E$ for each realization of the parameters from the 0d model as described in Methods. Then, the prior mean and covariance of $E$ are calculated from this set of $\{ E^{(i)}_L \}_{i=1}^{N_{MC}}$ realizations as

\begin{align}
    \mu_L (\bx) &= \frac{1}{N_{MC}} \sum_{m=1}^{N_{MC}} E^{(m)}_L (\bx) \\  k_L(\bx, \bx') &= \frac{1}{N_{MC}-1} \sum_{m=1}^{N_{MC}} (E^{(m)}_L(\bx)-\mu_L(\bx))(E^{(m)}_L(\bx')-\mu_L(\bx')).
\end{align}
Here, the superscript $L$ signifies that the corresponding  quantities are obtained from a numerical model that is based on approximation and has a lower fidelity than the experimental data. The estimates of $E$ in terms of the low-fidelity prior statistics are given as:

\begin{equation}\label{lf_mean}
    \mu_p(\bx^*) = \mu_L(\bxs) + \bc_L(\bxs)^T\bC^{-1}_L(\mathbf{E} - \mathbf{\boldsymbol{\mu}}_L) 
\end{equation}
\begin{equation}\label{lf_variance}
   \sigma^2_p(\bxs) = \sigma^2_L(\bxs) - \bc_L(\bxs)^T\bC^{-1}_L\bc_L(\bxs),
\end{equation}
where the components of $\bc_L(\bxs)$ and $\bC_L$ are given by $k_L$. 

When the physics-based model of a considered system is accurate the prior statistics computed from Eqs (\ref{lf_mean}) and (\ref{lf_variance}) were shown to provide an accurate model of the system given the system's measurements \cite{tartakovsky2019physicsHICSS,yang2018}. However, if the physics-based model cannot capture certain features of the system (as is the case with the 0d model), then the prior statistics computed would produce a poor GPR model of the system \cite{yang2019}. In the following section we demonstrate how the discrepancy between the observation data and the 0d model can be accounted for in the GPR prediction.

\subsection{Correcting bias in the 0d model}\label{cophik}

The systematic errors in the 0d model (also called model bias or misspecification) can be corrected by establishing the correlation between the 0d model predictions and the experimental data using the
CoKriging method \cite{yang2019,kennedy2000predicting}. This idea was originally formulated for systems where the data are sparse, so measurements of other states or other parameters are additionally used to model the data \cite{stein1991universal, knotters1995comparison}.
CoKriging has recently been used with success for data-driven  multifidelity modeling to predict system responses using data with two levels of fidelity \cite{legratiet2014, perdikaris2015multi} adapted from Kennedy and O’Hagan’s CoKriging framework \cite{kennedy2000predicting}.

 In the multifidelity framework, we treat $E$ observed in the experiments as high-fidelity data. We then use the 0d model to generate the low-fidelity data $\mathbf{E}_L = \left(E_{L}^{(1)}, \ldots,E_{L}^{(N_L)}\right)^T$ at locations $\mathbf{X}_L = \{ {\mathbf{x}^{(i)}_L} \}_{i=1}^{N_L} $, 
where 
$E_L^{(i)} \in \bbR$, 
 and 
$\bx_L^{(i)} \in \bbD \subseteq \bbR^d$. Note that in this work we generate $E_L^{(i)}$ using MC simulations (see  Section \ref{low_fidelity}) and $N_L = N_{MC}$, the number of realizations in the MC simulations. 

Following \cite{yang2019,kennedy2000predicting}, we define an auto-regressive model for the high-fidelity data, $E(\bx) = \rho E_L(\bx)+ E_D(\bx)$, where $E_L(\bx)$ is the GP model of the low-fidelity data, $E_D(\bx)$ is the GP model describing the mismatch between the high and low fidelity models, and $\rho$ is the regression parameter. The observation covariance matrix for this model is

\begin{equation}
    \tilde \bC = \begin{bmatrix}
    \bC_L(\bX_L, \bX_L) & \rho \bC_L(\bX_L, \bX) \\
    \rho \bC_L(\bX, \bX_L) & \rho^2 \bC_L(\bX, \bX) + \bC_D(\bX, \bX) 
    \end{bmatrix}
\end{equation}
where the elements of the $\bC_L$ matrices are computed as $k_L(\mathbf{x},\mathbf{x}')$, where $k_L$ is the covariance function of $E_L$ and
the elements of the $\bC_D$ matrix are given by $k_D(\mathbf{x},\mathbf{x}')$, where $k_D$ is the covariance function of $E_D$.

The posterior mean and variance of $E$  at any point $\bx^* \in \bbD$ conditioned on the measurements  of  $\mathbf{E}$ and  $\mathbf{E}_L$ are given by

\begin{align} \label{cophik_mean}
    \mu_p(\bx^*) &= \mu_{H}(\bx^*) + \tilde \bc(\bx^*)^T \tilde \bC^{-1}\left(\tilde{\mathbf{E}} - \Tilde{\boldsymbol{\mu}}\right) \\ \label{cophik_var}
    \sigma^2_p (\bx^*) &= \rho^2\hat \sigma_L^2(\bx^*) + \sigma_D^2(\bx^*)-\tilde \bc(\bx^*)^T\tilde \bC^{-1}\tilde \bc(\bx^*),
\end{align}
where 
 $\tilde{\mathbf{E}} = (\mathbf{E}_L^T, \mathbf{E}^T)^T$, 
$\mu(\bx^*) = \rho\mu_L(\bx^*)+\mu_D(\bx^*)$, $\sigma_L^2(\bx^*) = k_L(\bx^*, \bx^*)$, $\sigma_D^2(\bx^*) = k_D(\bx^*, \bx^*)$,  $\Tilde{\boldsymbol{\mu}} = (\boldsymbol{\mu}_L^T, \boldsymbol{\mu}^T)^T$, $\tilde \bc(\bx^*) = (\rho \bc_L(\bx^*)^T, \bc(\bx^*)^T)^T$, and 
$k(\bx, \bx') = \rho^2k_L(\bx, \bx')+k_D(\bx, \bx').$

In the data-driven approaches of \cite{kennedy2000predicting,stein1991universal, knotters1995comparison,legratiet2014, perdikaris2015multi}, the prior statistics (i.e., $\mu_L(\bx^*)$, $\mu_D(\bx^*)$, $k_L(\bx, \bx')$, and $k_D(\bx, \bx')$) are learned by maximizing the likelihood function of $Y$. Here, we compute $\mu_L(\bx^*)$ and $k_L(\bx, \bx')$ from the Monte Carlo simulations based on the 0d model with random parameters $k_{1}, k_2, \sigma_e$, and $S$ as described in Section \ref{phik}. Then, we compute $\mu_L(\bx^*)$ and $k_L(\bx, \bx')$ using the data set $\mathbf{E_D} = \mathbf{E} - \rho \mathbf{E}_L^{X}$ with the corresponding locations $\mathbf{X}$  (where  $\mathbf{E}_L^{X} = [E_{L,1}, E_{L,1},...,E_{L,N}]^T$ and $E_{L,i} = \mu_L(\mathbf{x}_i)$ ($\mathbf{x}_i \in \mathbf{X}$)) by minimizing the log likelihood:

\begin{equation}\label{likelihooh}
\ln L(\mu_D, \sigma_D, \lambda_D) 
=\frac12 (\mathbf{E}_D - \boldsymbol{\mu}_D)^T \mathbf{C}^{-1}_D (\mathbf{E}_D - \boldsymbol{\mu}_D)
-\frac12 \ln |\mathbf{C}_D| - \frac{N}{2} \ln 2\pi.
\end{equation}
Here, we set $\rho = 1.0$ and assume that $E_D(\mathbf{x})$ has the prior exponential covariance function

\begin{equation}
    k_D(\mathbf{x},\mathbf{x}') = \sigma_D^2 \exp \left( - \frac{|\mathbf{x} - \mathbf{x}'|}{\lambda_D} \right),  
\end{equation}
and constant $\mu_D$. In Eq (\ref{likelihooh}), the $(i,j)$ component of the covariance matrix $\mathbf{C}_D$ is given by $k_D(\mathbf{x}_i,\mathbf{x}_j)$ ($(\mathbf{x}_i,\mathbf{x}_j) \in \mathbf{X}_D$).

\subsection{Determining the SOC scaling parameter}

\begin{figure}[h]
\centering
\includegraphics[width=0.6\linewidth]{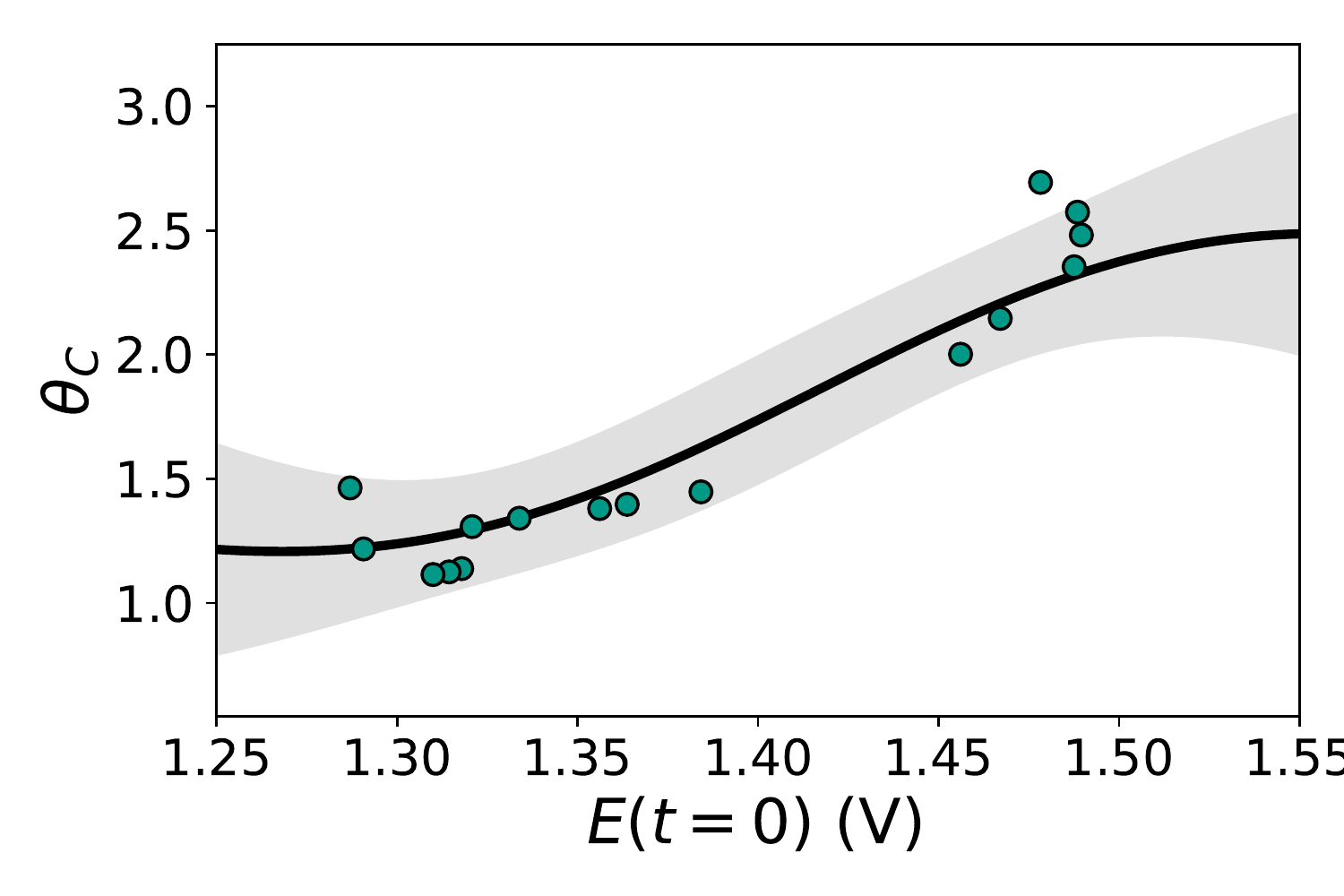}
\caption{GPR model prediction of $\theta_C$ as a function of $E_0=E(t=0)$ (black line) trained on data from 16 experiments (circles). The shaded area corresponds to the GPR model  uncertainty.}\label{fig:scaling_param}
\end{figure}

A key step in the data preparation for training the CoKriging model is to rescale SOC in different experiments according to Eqs. (\ref{eq:SOC_scale1}) and (\ref{eq:SOC_scale2}). For data from an experiment $i$, the scaling parameters in Eqs. (\ref{eq:SOC_scale1}) and (\ref{eq:SOC_scale2}) (i.e., $SOC_i(0)$ and $\max(SOC_i)$) are given by the experimental data. Furthermore, for a given set of input parameters, the CoKriging model predicts voltage as a function of the scaled SOC. However, to covert the scaled SOC to time, one needs to evaluate scaling parameters as functions of the input parameters. In the following, we introduce two scaling parameters denoted by $\theta_C = 1/(\max(SOC_C)-SOC_C(0))$ and $\theta_D = 1/(\max(SOC_D)-SOC_D(0))$. When $\max(SOC_C)$ and $\max(SOC_D)$ are known these scaling parameters are trivial to find. However,  we may not know $SOC_C(0)$ and $SOC_D(0)$ for arbitrary conditions, so we first need to estimate them or, equivalently, the scaling parameters. 
We compute $\theta_C$ using the standard GPR model trained on a data set consisting of the scaling parameters $\theta_C$  and corresponding  initial voltages $E_0=E(t=0)$ from the known experiments. We note that in the considered lab experiments, $-\theta_D = 0.05+\theta_C$, which allows us to estimate $\theta_D$ given $\theta_C$ without needing the voltage at any SOC during the discharge cycle. 
 We assumed that there is uncertainty in the $\theta_C$ values obtained from the experiments with the variance $\sigma^2_\theta = 0.5$ The non-zero $\sigma^2_\theta$ reflects the fact that  the different values of $\theta_C$ were obtained from the experiments with very similar values of $E_0$. The positive value of $\sigma^2_\theta$ is also needed to make the measurement covariance matrix well-conditioned.  Figure \ref{fig:scaling_param} shows the GPR model prediction of $\theta_C$ as a function of $E_0$ and the prediction  uncertainty. The GPR model uncertainty increase outside of the range $1.3-1.5$V of $E_0$, where the data are available.
This uncertainty can be reduced by conducting experiments with the expended range of $E_0$. Within $1.3-1.5$V range, the uncertainty in GPR model prediction can be reduced by considering additional variables that affect $\theta_c$.

\section{The multifidelity model prediction of charge and discharge curves}
\label{sec:results}

\begin{figure}[ht]
\begin{subfigure}{\textwidth}
\centering
\includegraphics[width=0.28\textwidth]{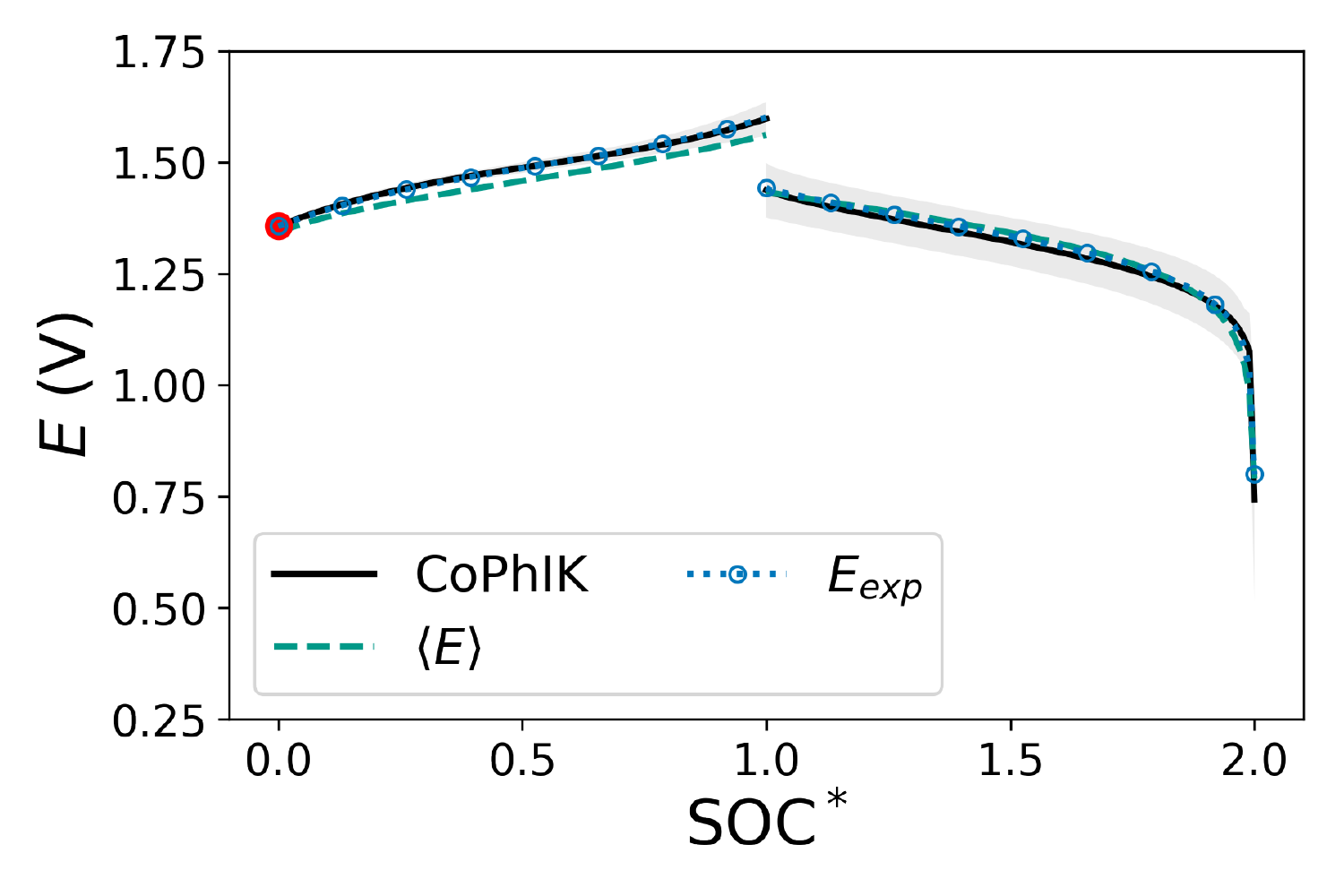}
\includegraphics[width=0.28\textwidth]{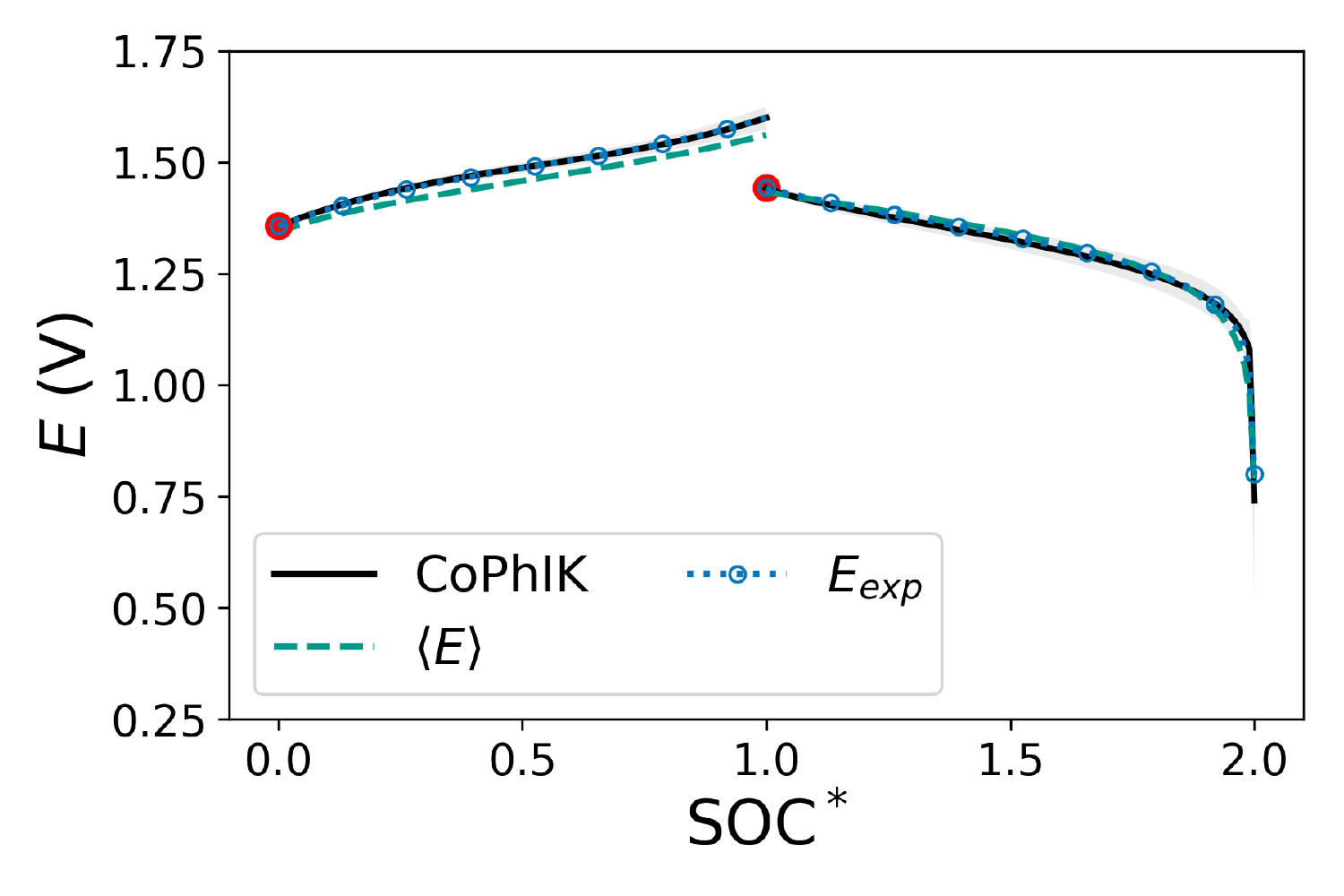}
\includegraphics[width=0.28\textwidth]{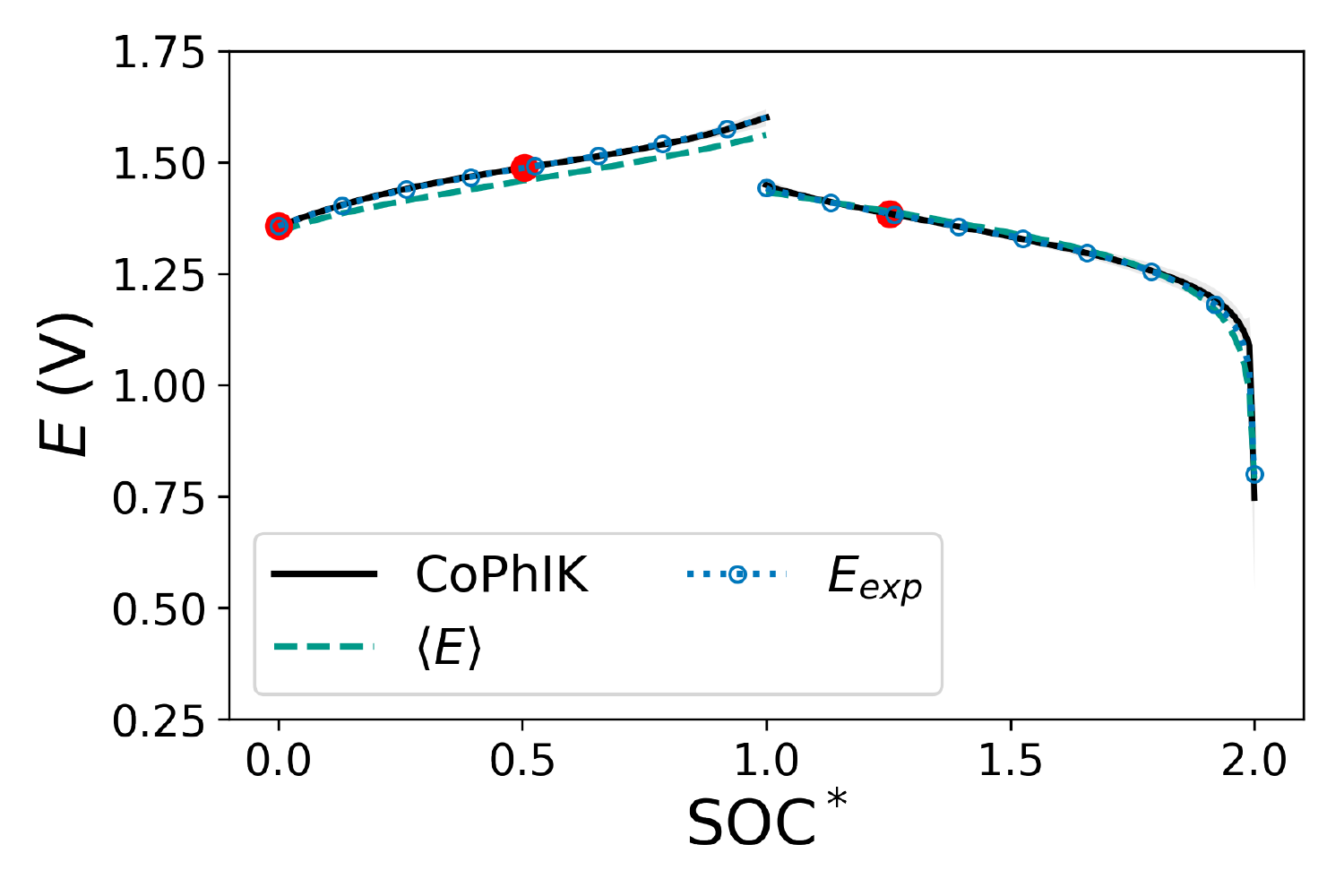}\\
\includegraphics[width=0.28\textwidth]{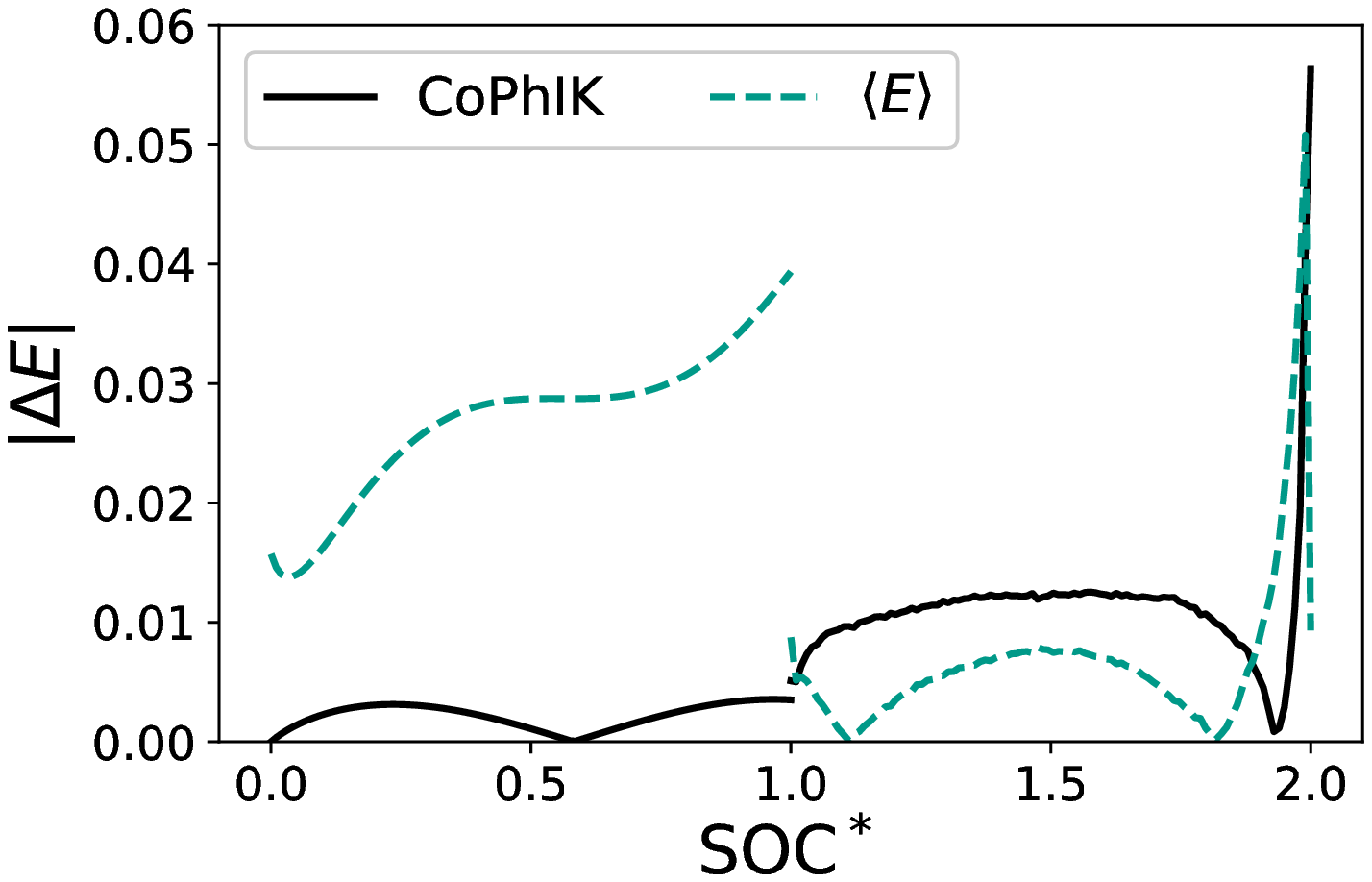}
\includegraphics[width=0.28\textwidth]{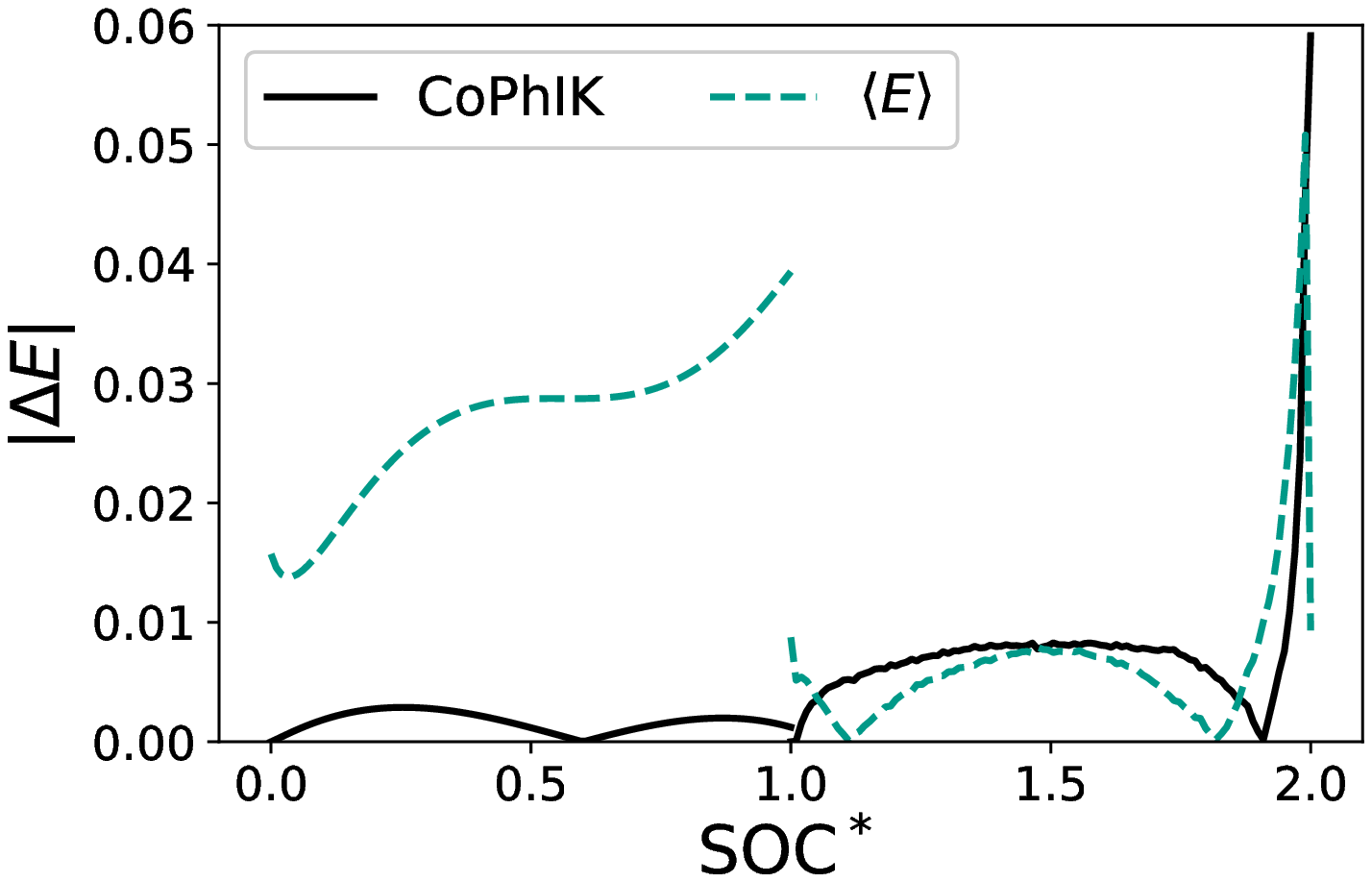}
\includegraphics[width=0.28\textwidth]{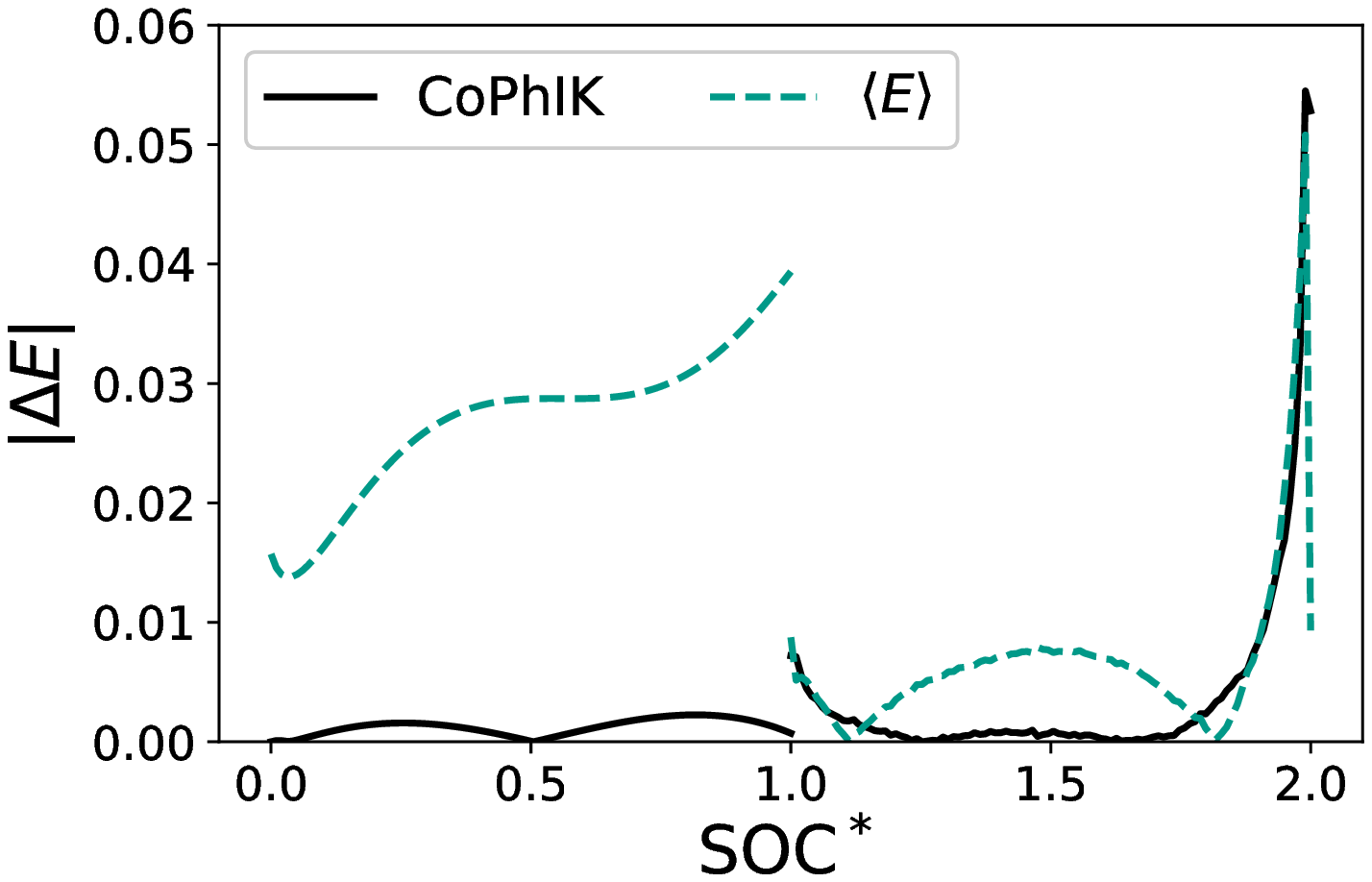}
\caption{}
\end{subfigure}
\begin{subfigure}{\textwidth}
\centering
\includegraphics[width=0.28\textwidth]{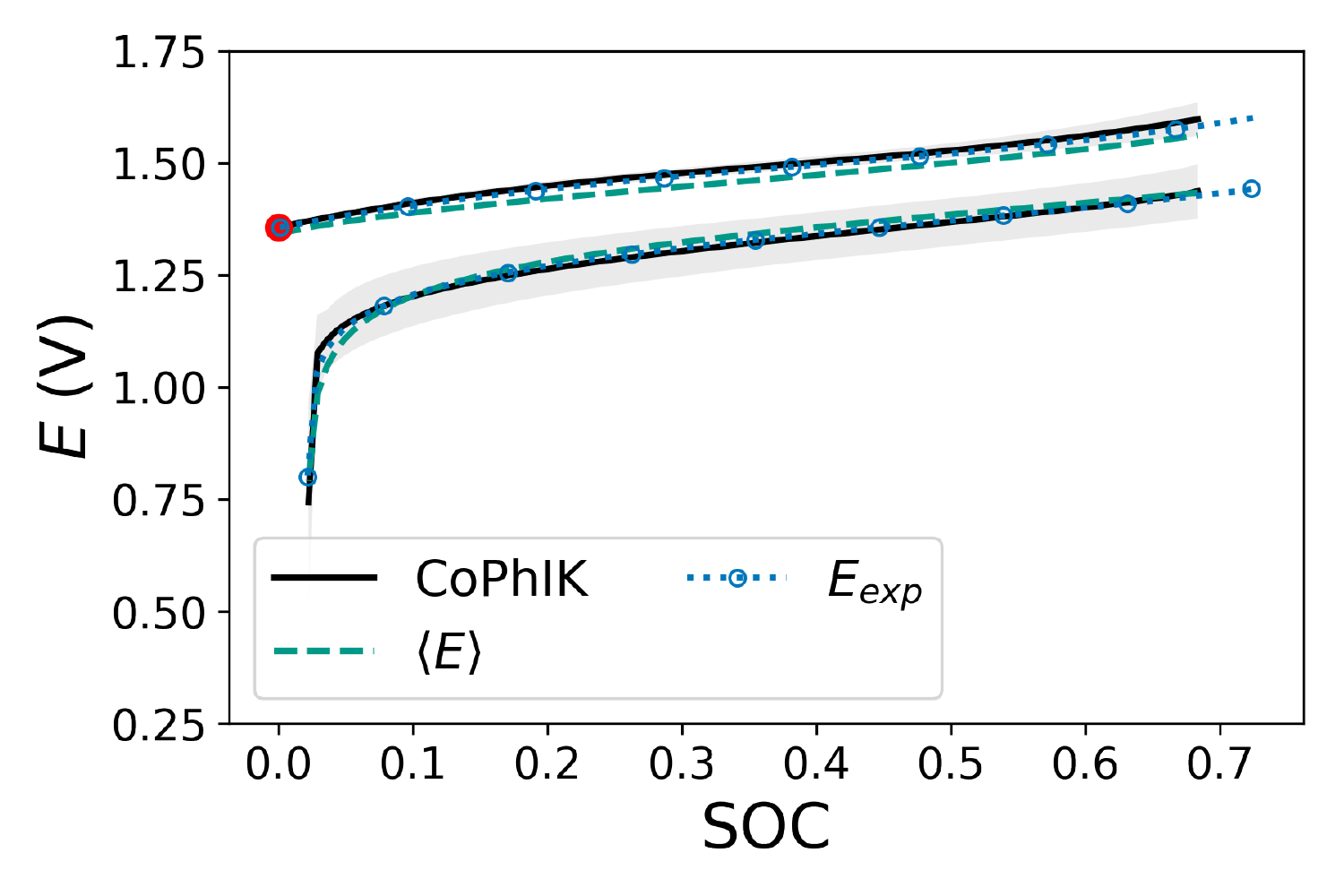}
\includegraphics[width=0.28\textwidth]{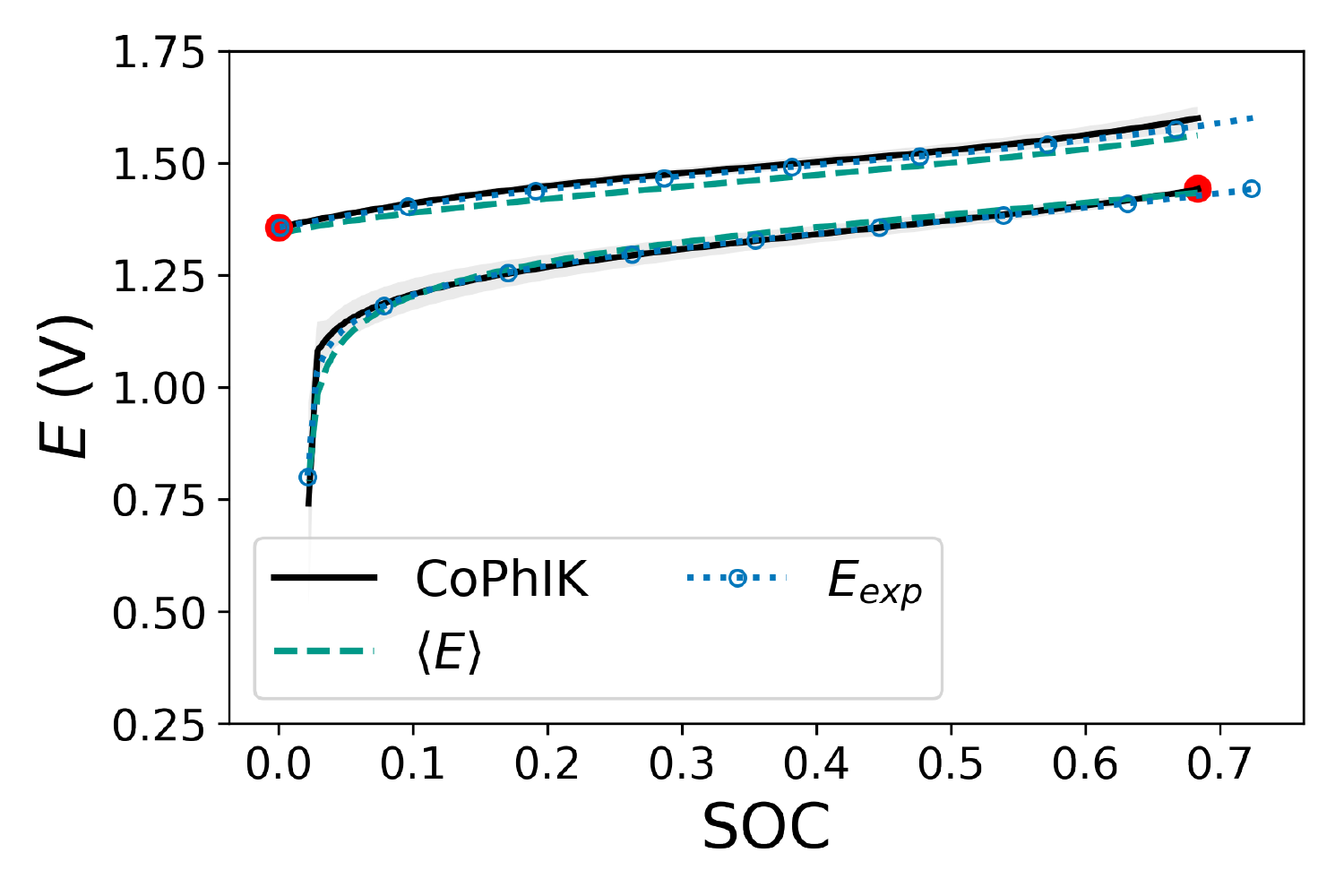}
\includegraphics[width=0.28\textwidth]{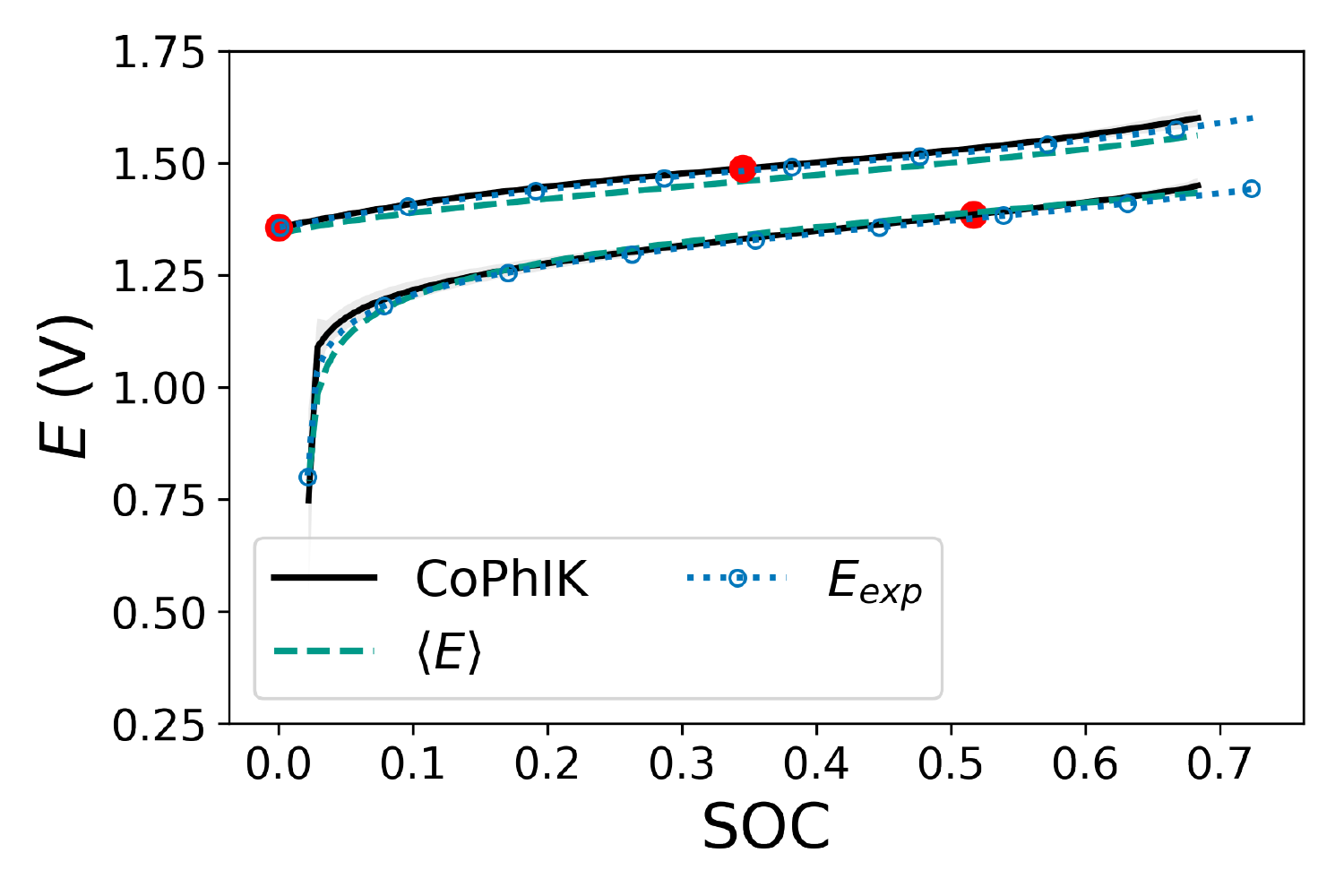}\\
\includegraphics[width=0.28\textwidth]{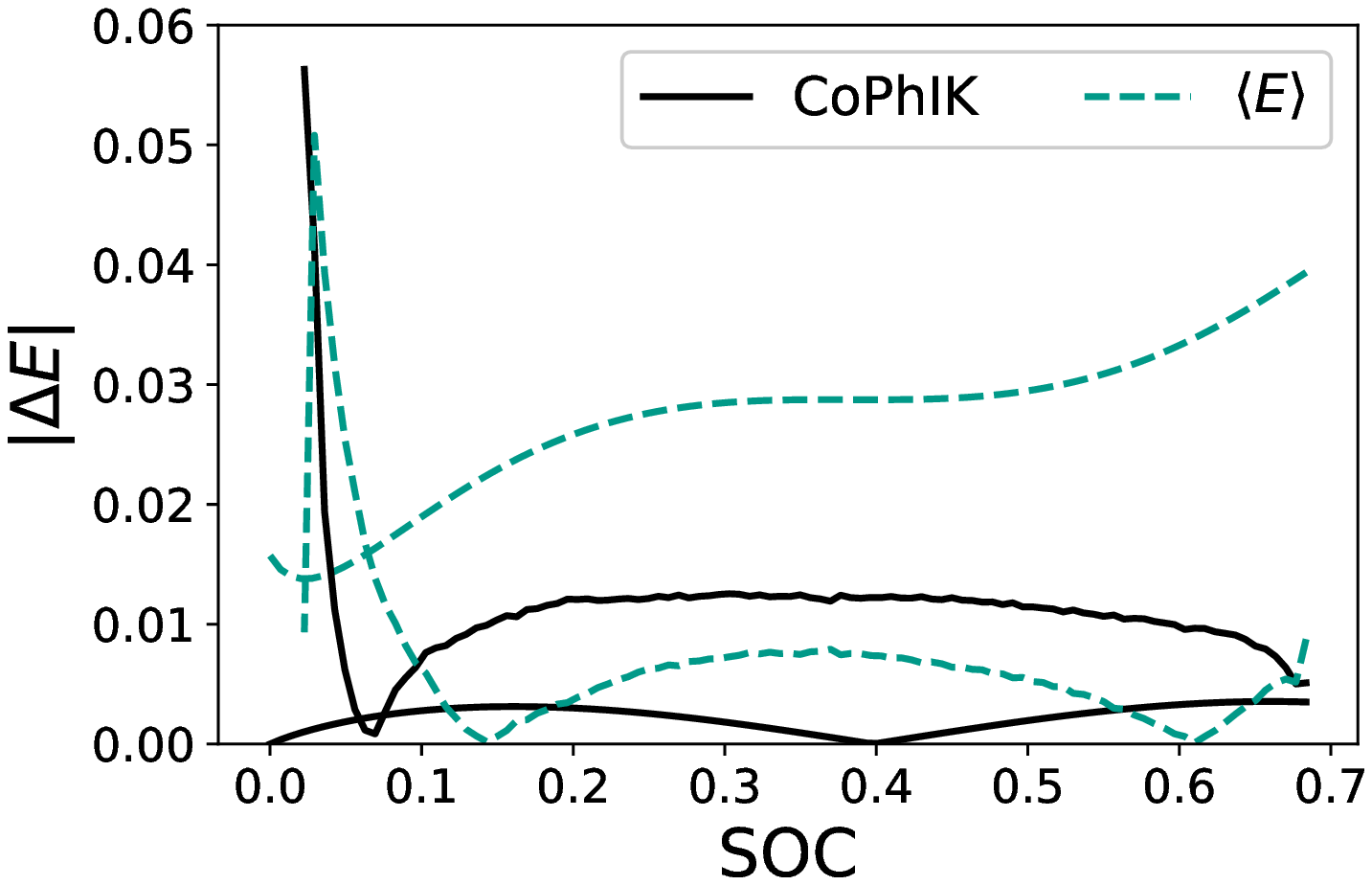}
\includegraphics[width=0.28\textwidth]{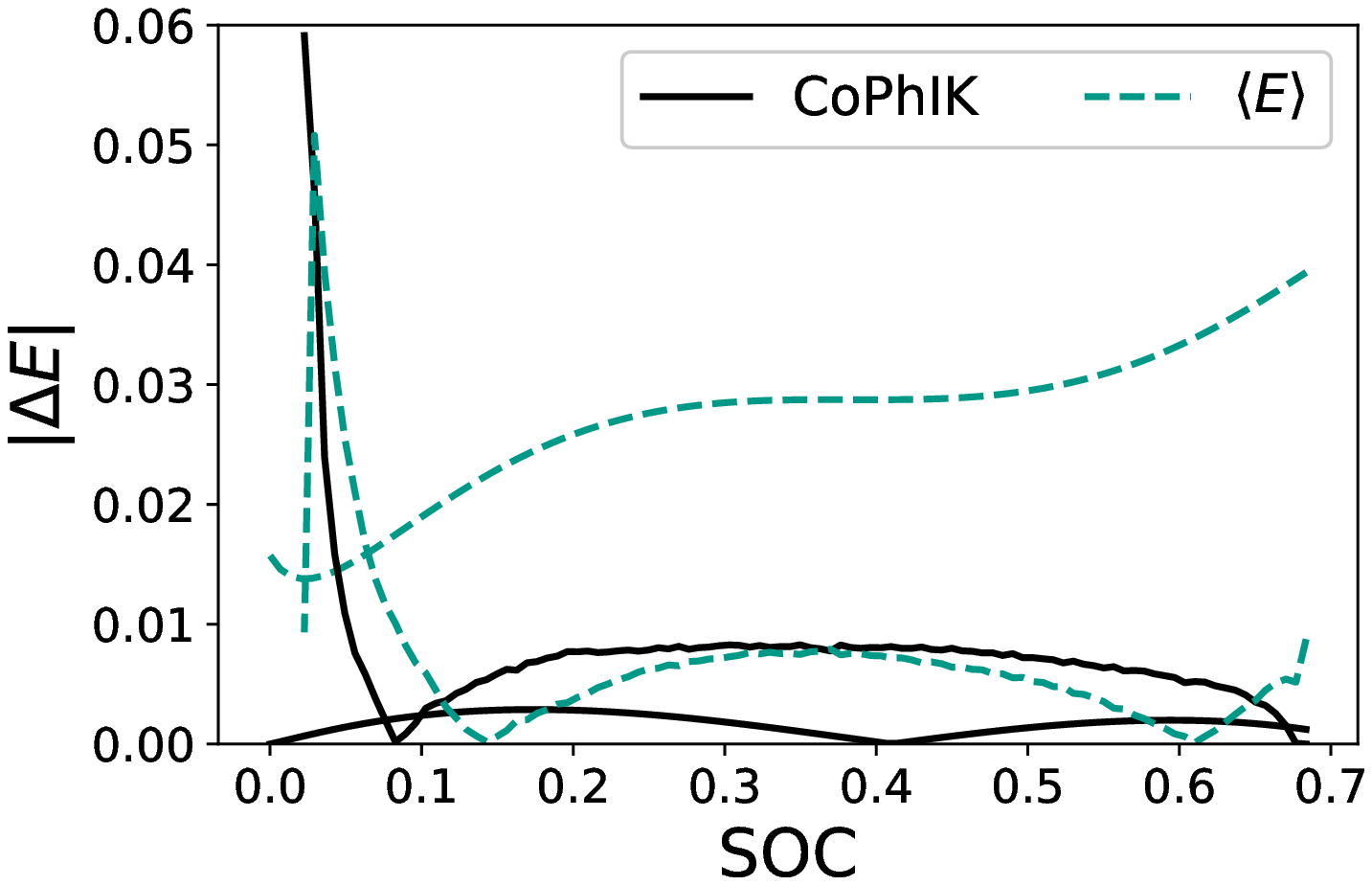}
\includegraphics[width=0.28\textwidth]{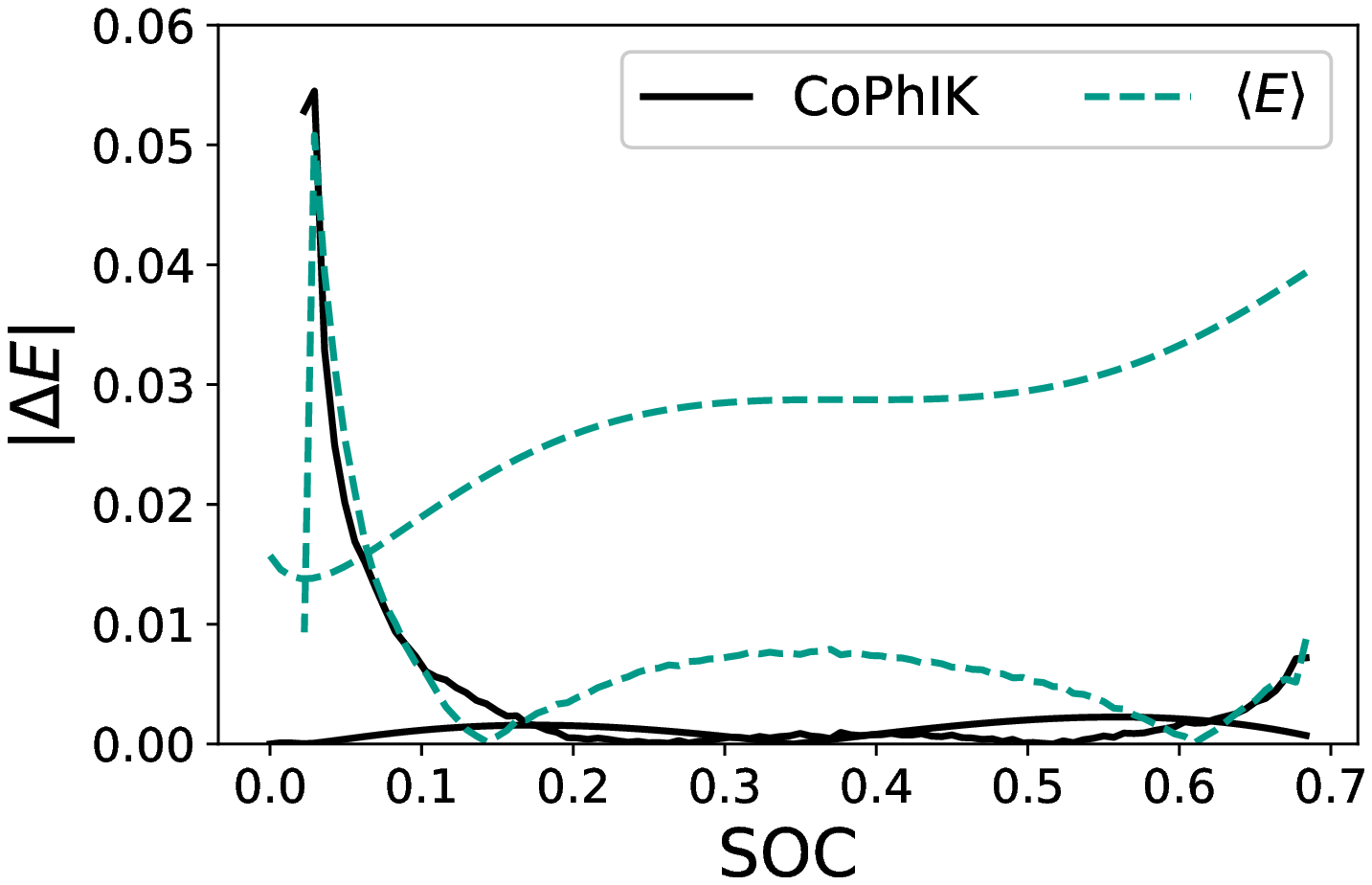}
\caption{}
\end{subfigure}
\caption{Example of CoPhIK results and relative errors for Experiment 2 with $1$, $2$, and $3$ data points included using (a) scaled and (b) unscaled SOC. The top row shows the resulting charge-discharge curve, the bottom row shows the relative errors. We plot the mean of the low-fidelity dataset ($\langle E \rangle$), the experimental data ($E_{exp}$), and the variance of the CoPhIK results ($E_{var}$).} \label{fig:CoPhIK1}
\end{figure}


CoPhIK can accurately predict the behavior of the charge-discharge curves, as shown in Fig. \ref{fig:CoPhIK1}. With only one data point used, $E(t=0)$, CoPhIK matches the experimental data well, and accurately captures the jump between the charge and discharge cycles. The largest disagreement between the experimental voltage and the CoPhIK results occurs at the tail of the discharge curve, where we have very few measurements available to accurately capture the curve. The variance in the CoPhIK results, shown as the shaded regions in Fig. \ref{fig:CoPhIK1}, decreases as more measurements are used in the CoPhIK prediction. 

We performed single-fidelity PhIK and multi-fidelity CoPhIK simulations for all sixteen experiments in our dataset and computed the average of the discrete $\ell_2$ and $\ell_\infty$ norms, as shown in table \ref{tab:errors}. The average $\ell_2$ error with CoPhIK is approximately two orders of magnitude lower than the average error with PhIK, and the CoPhIK $\ell_\infty$ error is approximately one order of magnitude smaller. By including the available experimental data we are able to reduce the error in our predictions over a fully physics-informed approach. 

\begin{figure}[h]
\centering
\includegraphics[width=0.3\linewidth]{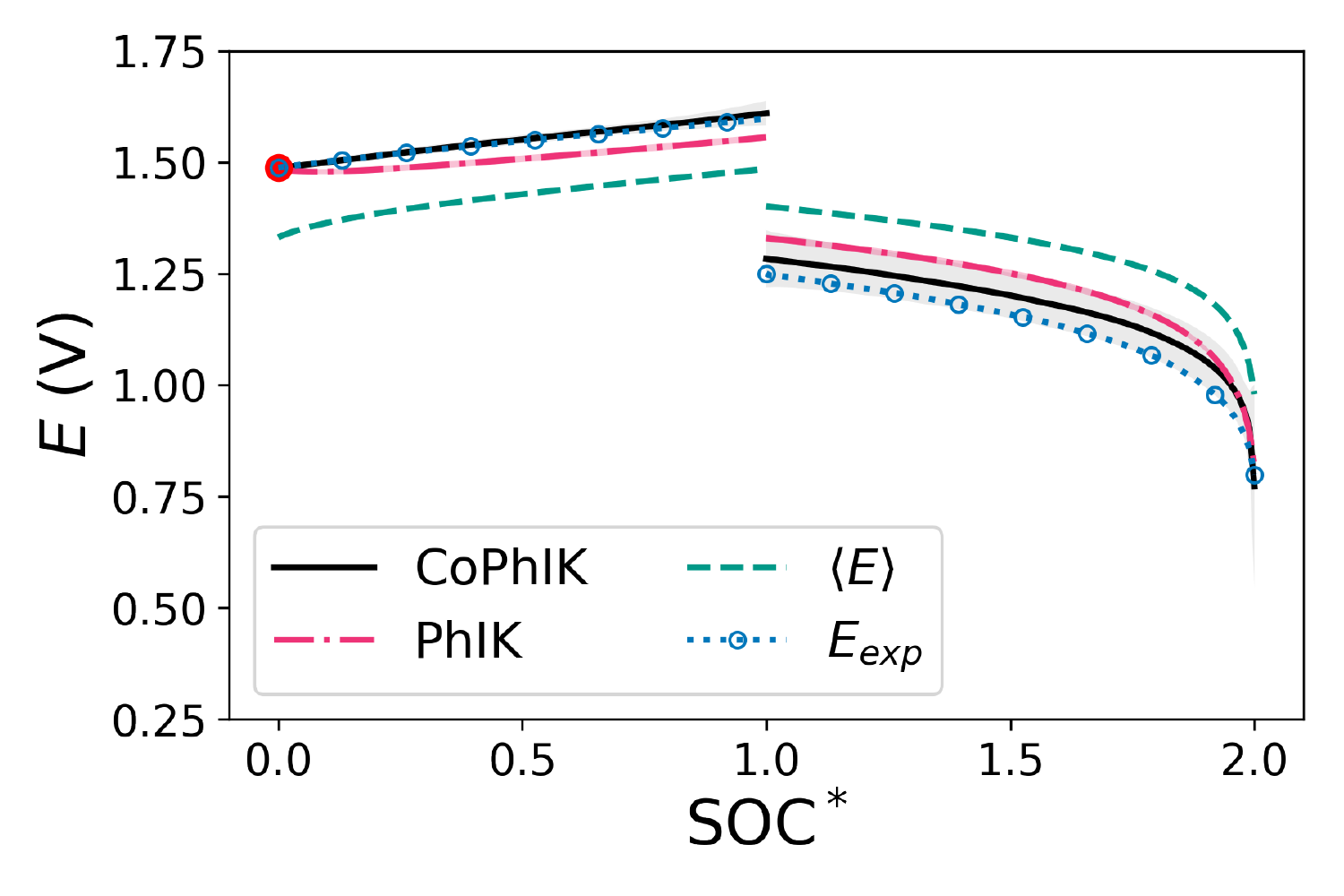}
\includegraphics[width=0.3\linewidth]{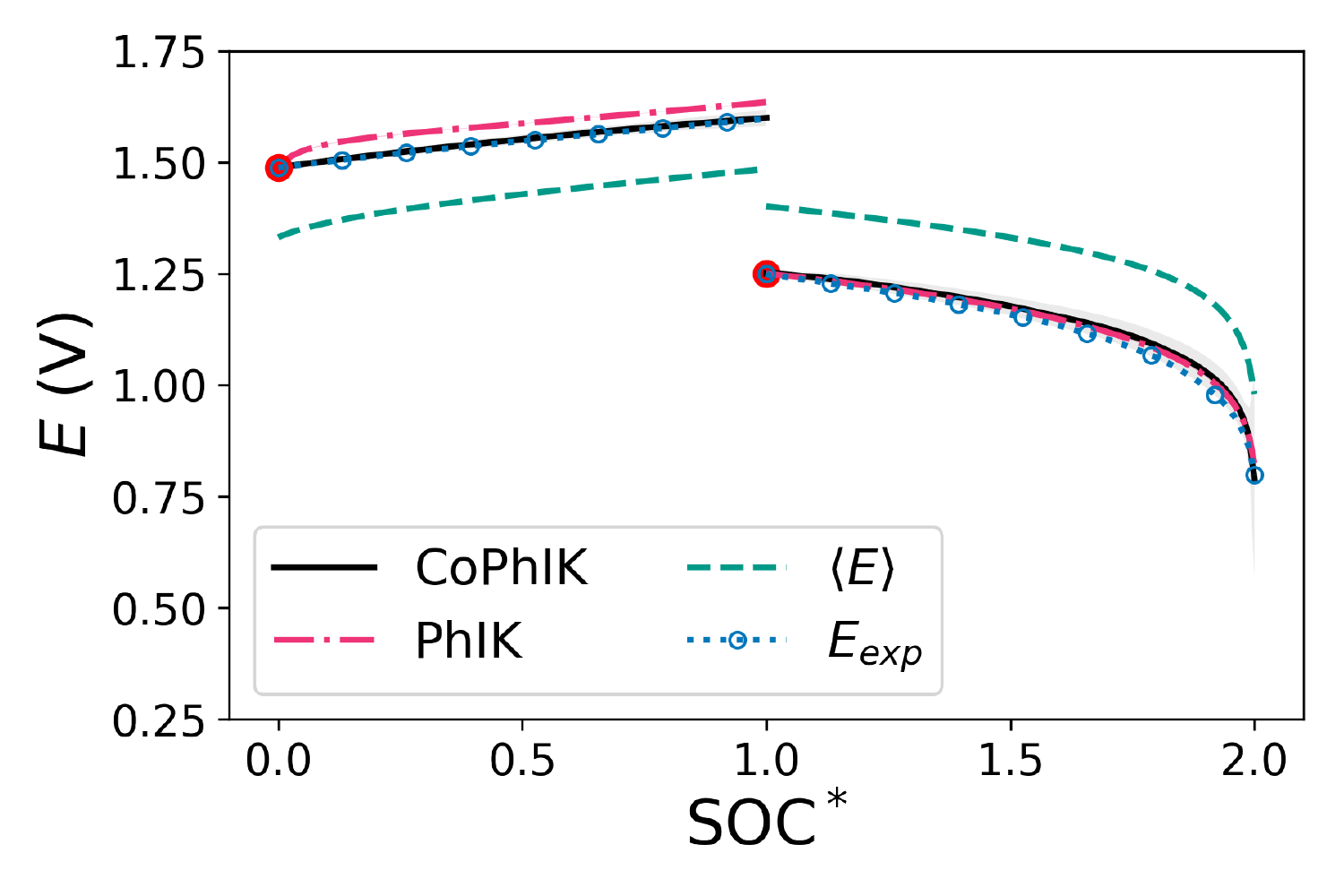}
\includegraphics[width=0.3\linewidth]{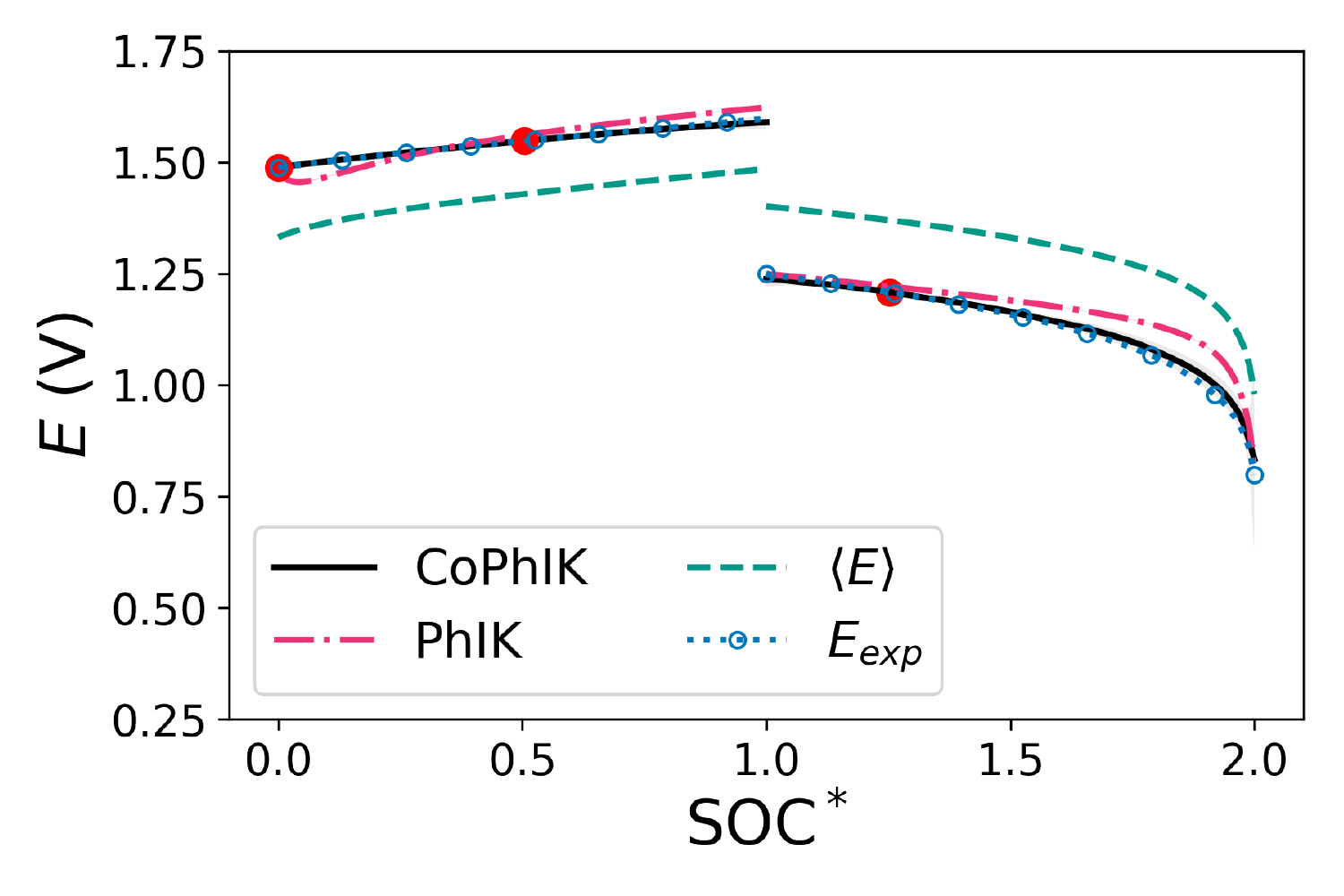}\\
\includegraphics[width=0.3\linewidth]{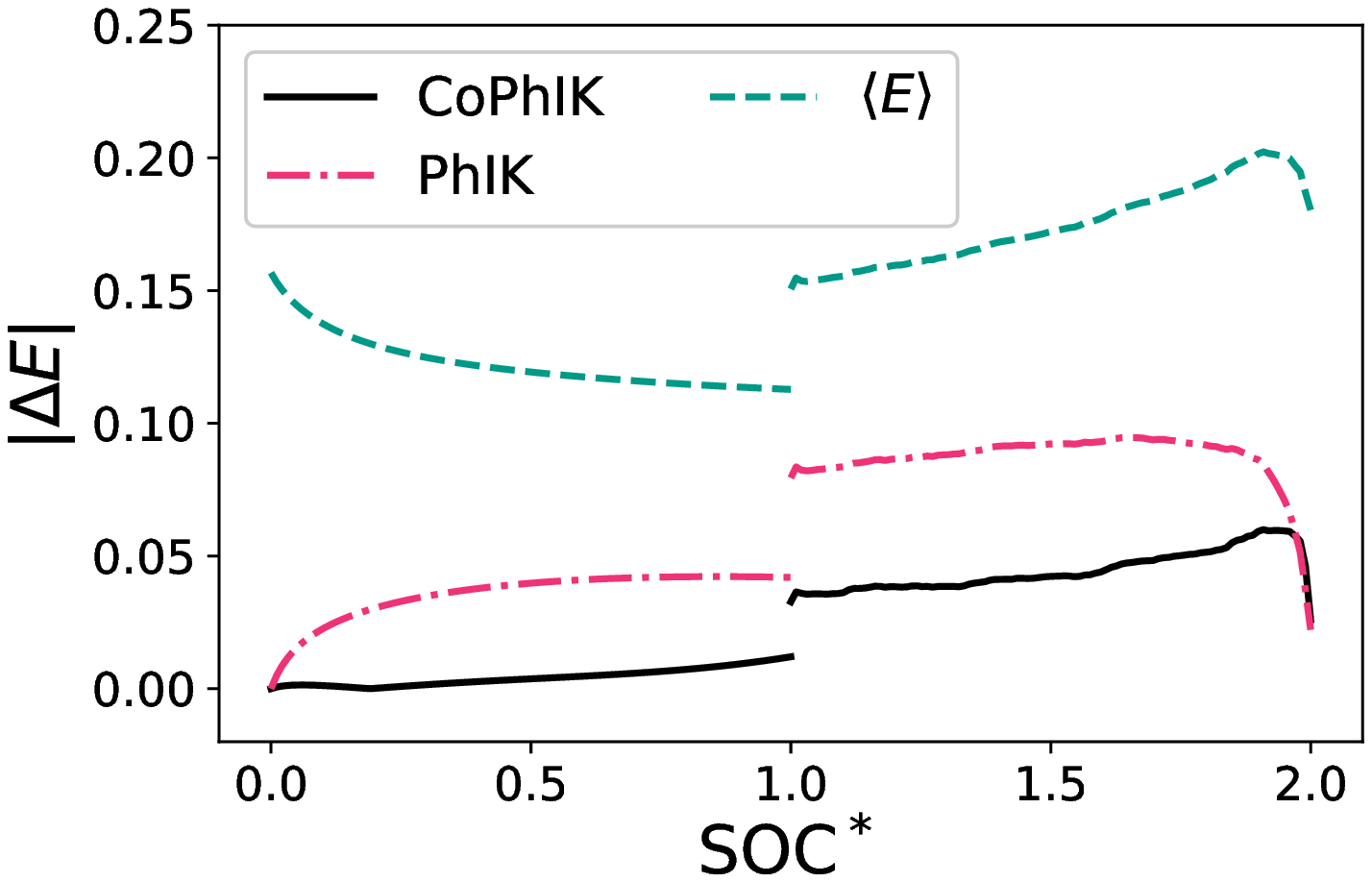}
\includegraphics[width=0.3\linewidth]{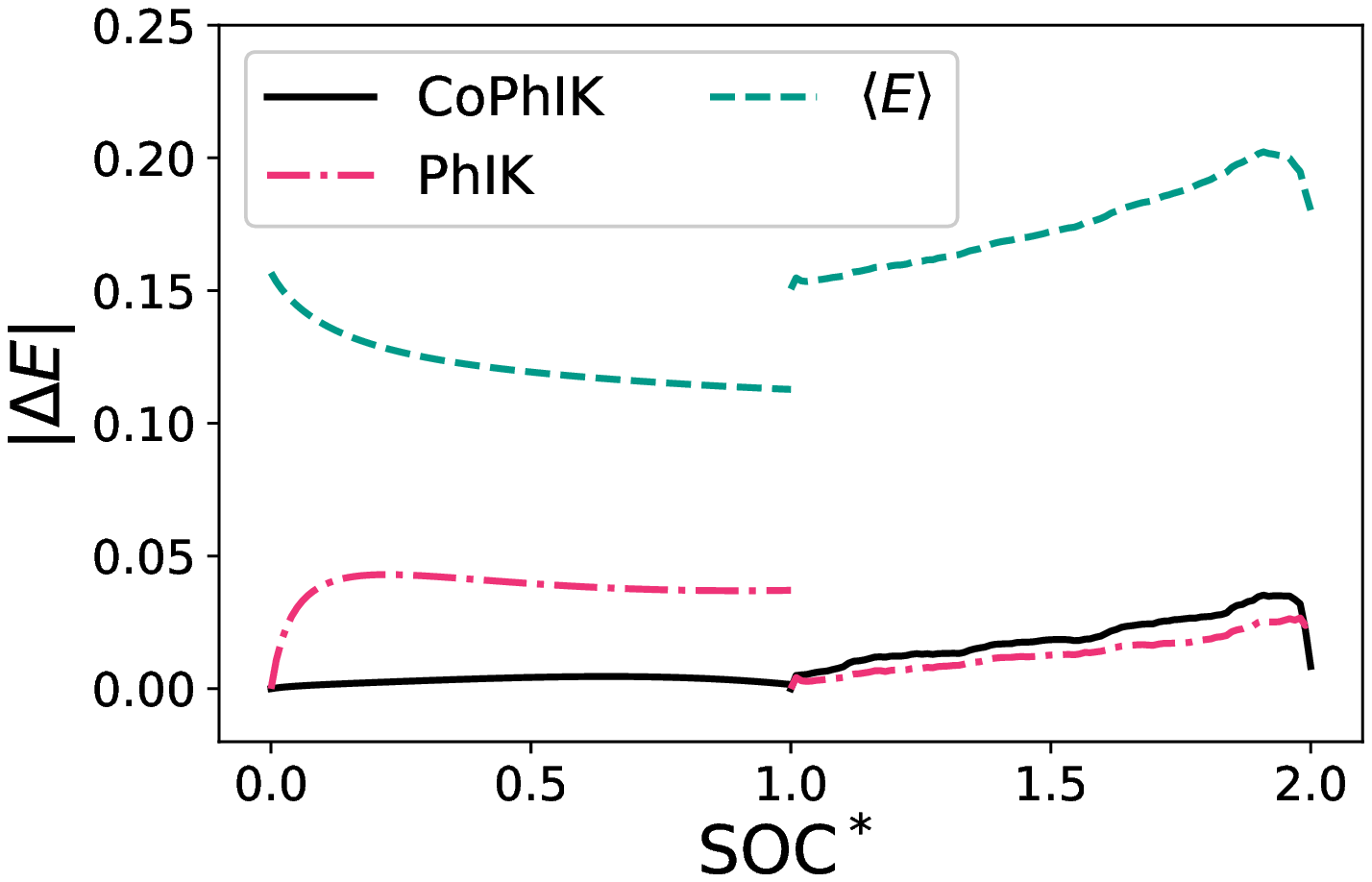}
\includegraphics[width=0.3\linewidth]{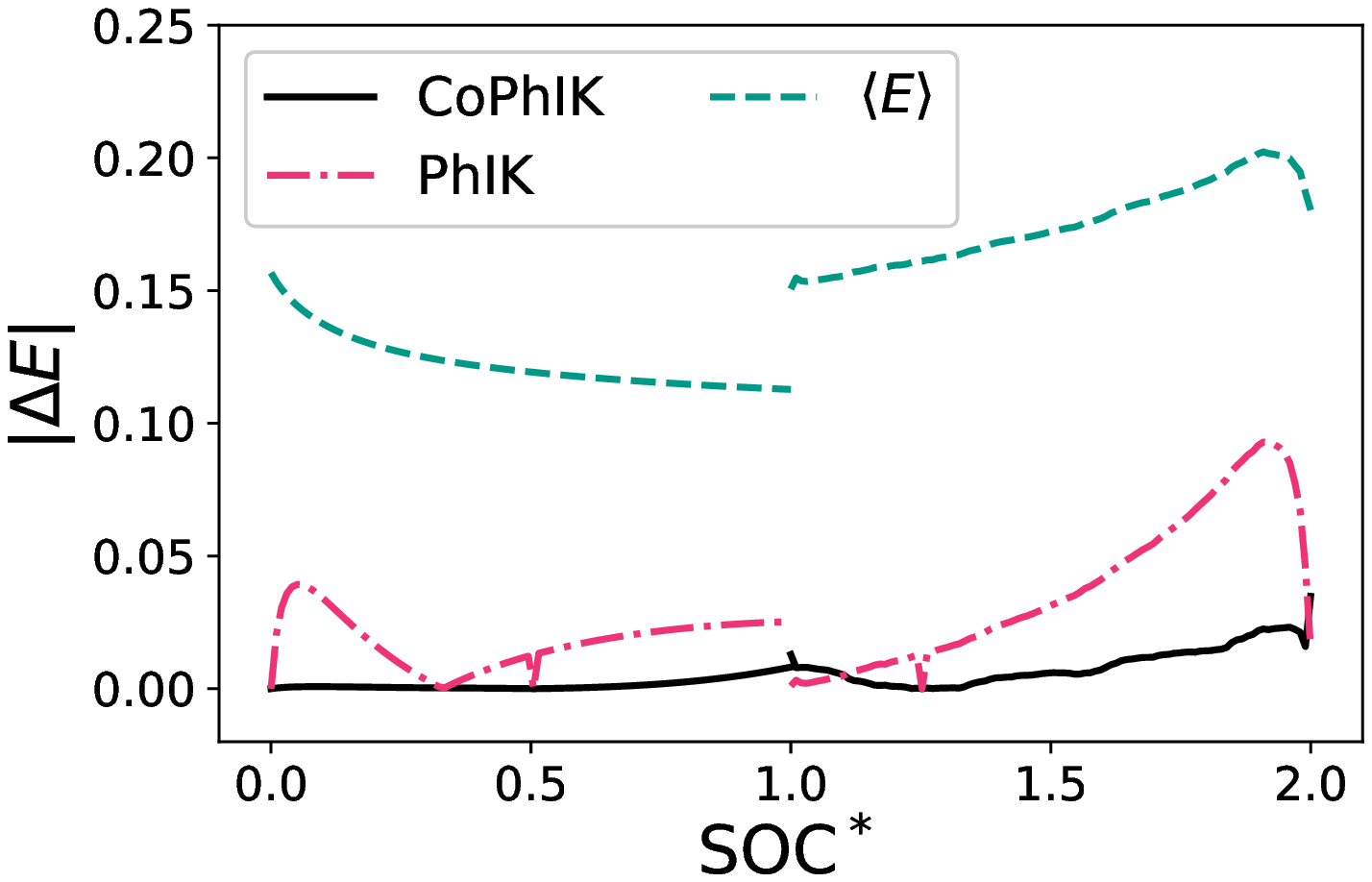}
\caption{Comparison between PhIK and CoPhIK results for Experiment 9. We plot the mean of the low-fidelity dataset ($\langle E \rangle$), the experimental data ($E_{exp}$), and the variances from CoPhIK (grey) and PhIK (pink). The top row shows the charge-discharge curve while the bottom row shows the relative errors.} \label{fig:CoPhIKcomp}
\end{figure}

In Fig. \ref{fig:CoPhIKcomp} we provide a comparison between CoPhIK and PhIK. It is clear that PhIK struggles to capture a physical shape for the charge and discharge curves, resulting in overall larger errors. PhIK shows a smaller variance than CoPhIK because the uncertainty of PhIK is bounded by the \emph{parametric} uncertainty in the OD model and does not take into account the \emph{model} uncertainty in the 0d model associated with the simplifying assumptions in the 0d model. 

\begin{table}
\centering
\begin{tabular}{ c| c c c | c c c} \hline
 & $\ell_2$, 1pt. & $\ell_2$, 2pts. & $\ell_2$, 3pts. & $\ell_\infty$, 1pt. &$\ell_\infty$, 2pts. &$\ell_\infty$, 3pts. \\ \hline
PhIK & $1.40\times 10^{-2}$ & $1.26\times 10^{-2}$ & $1.28\times 10^{-2}$ & $9.56\times 10^{-1}$ & $9.50\times 10^{-1}$ & $9.31\times 10^{-1}$  \\
CoPhIK &$4.36\times 10^{-4}$ & $1.74\times 10^{-4}$ & $1.11\times 10^{-4}$ &  $8.81\times 10^{-2}$ & $8.38\times 10^{-2}$ & $8.69\times 10^{-2}$ 
\\ \hline
\end{tabular} 
\caption{Average of the $\ell_2$ and $\ell_\infty$ errors between the experimental data with PhIK and CoPhIK.} \label{tab:errors}
\end{table}

\subsection{Sensitivity to choice of initial parameters}\label{sensitivity}

\begin{figure}[h]
\centering
\includegraphics[width=0.3\linewidth]{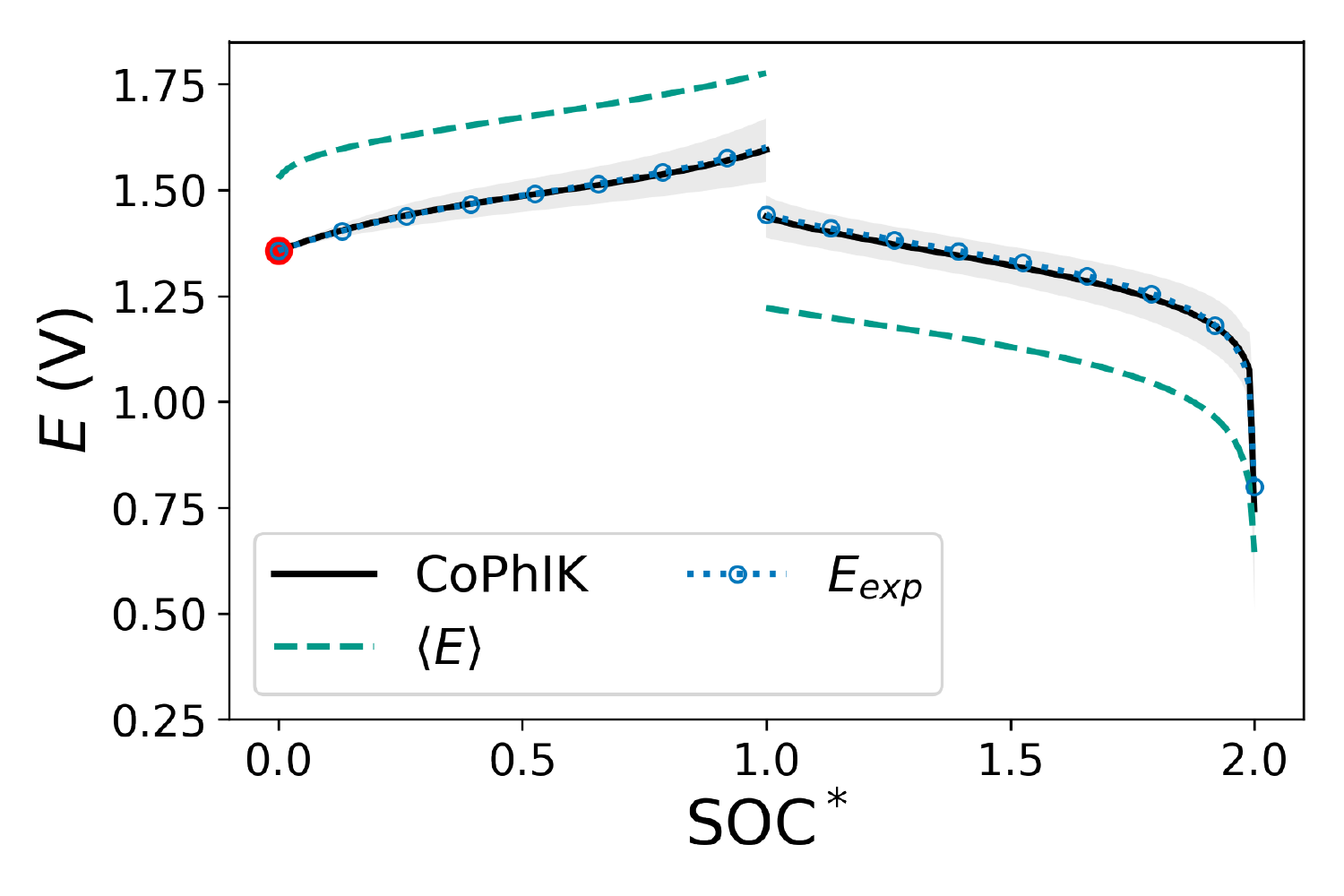}
\includegraphics[width=0.3\linewidth]{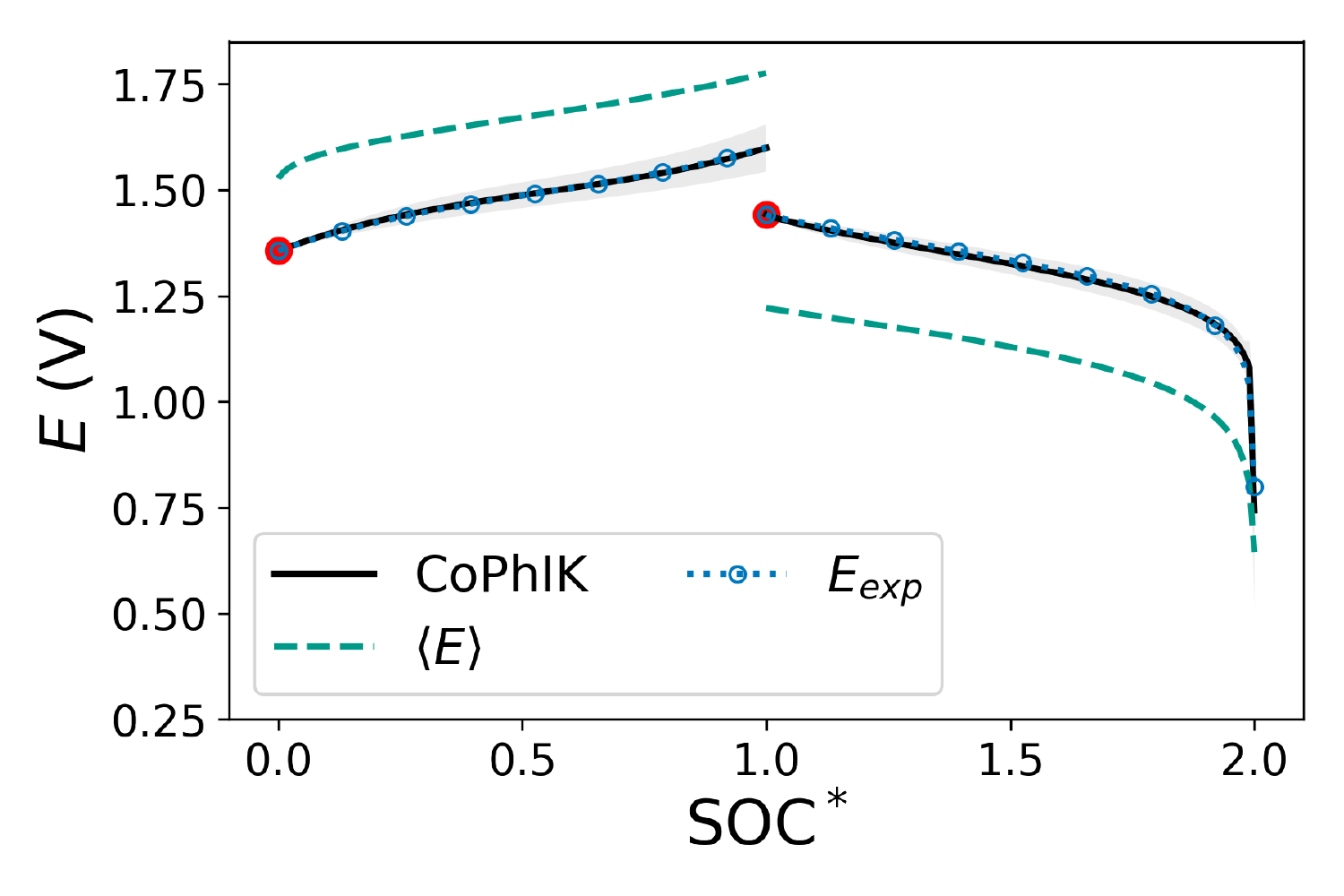}
\includegraphics[width=0.3\linewidth]{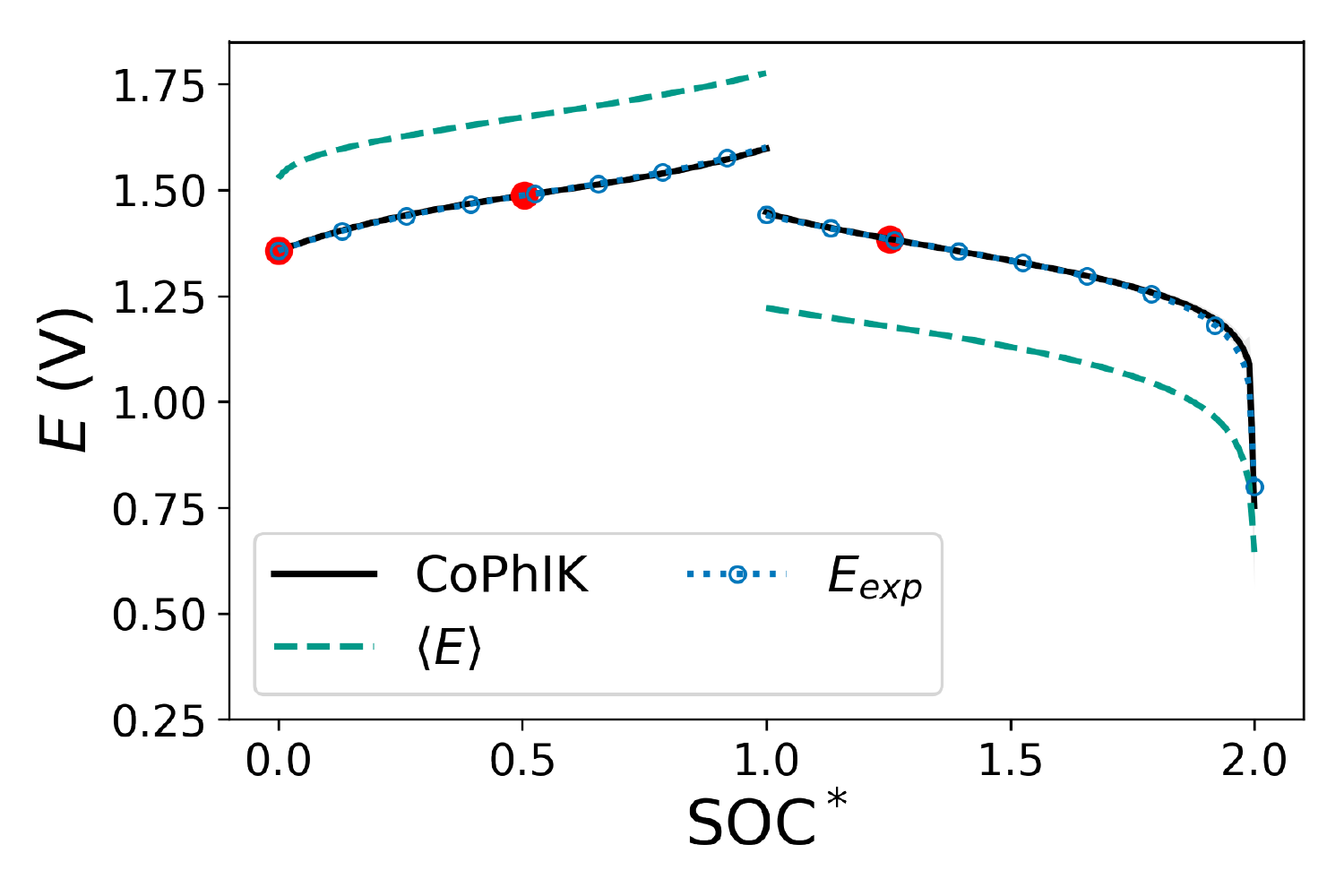}\\
\includegraphics[width=0.3\linewidth]{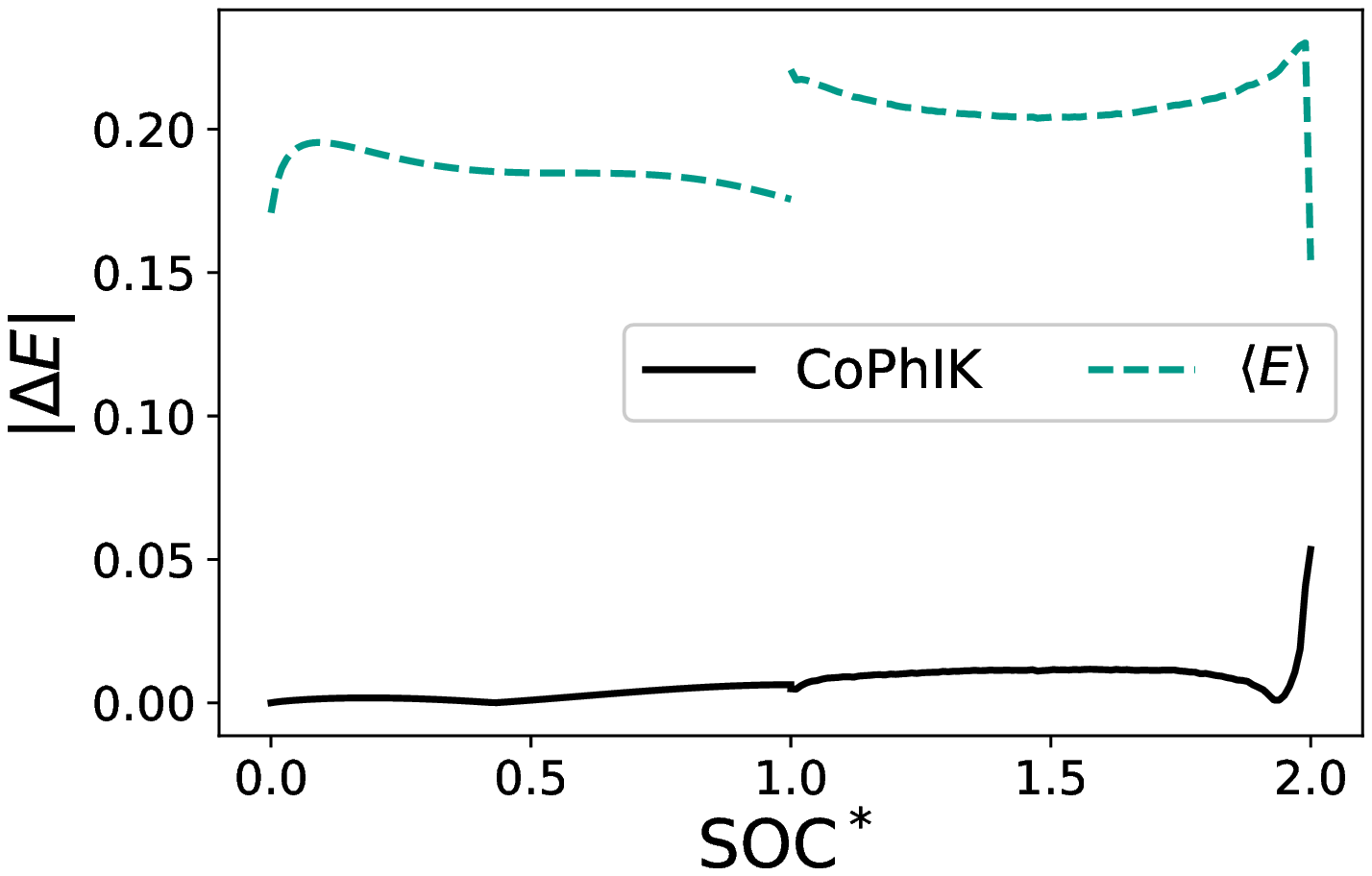}
\includegraphics[width=0.3\linewidth]{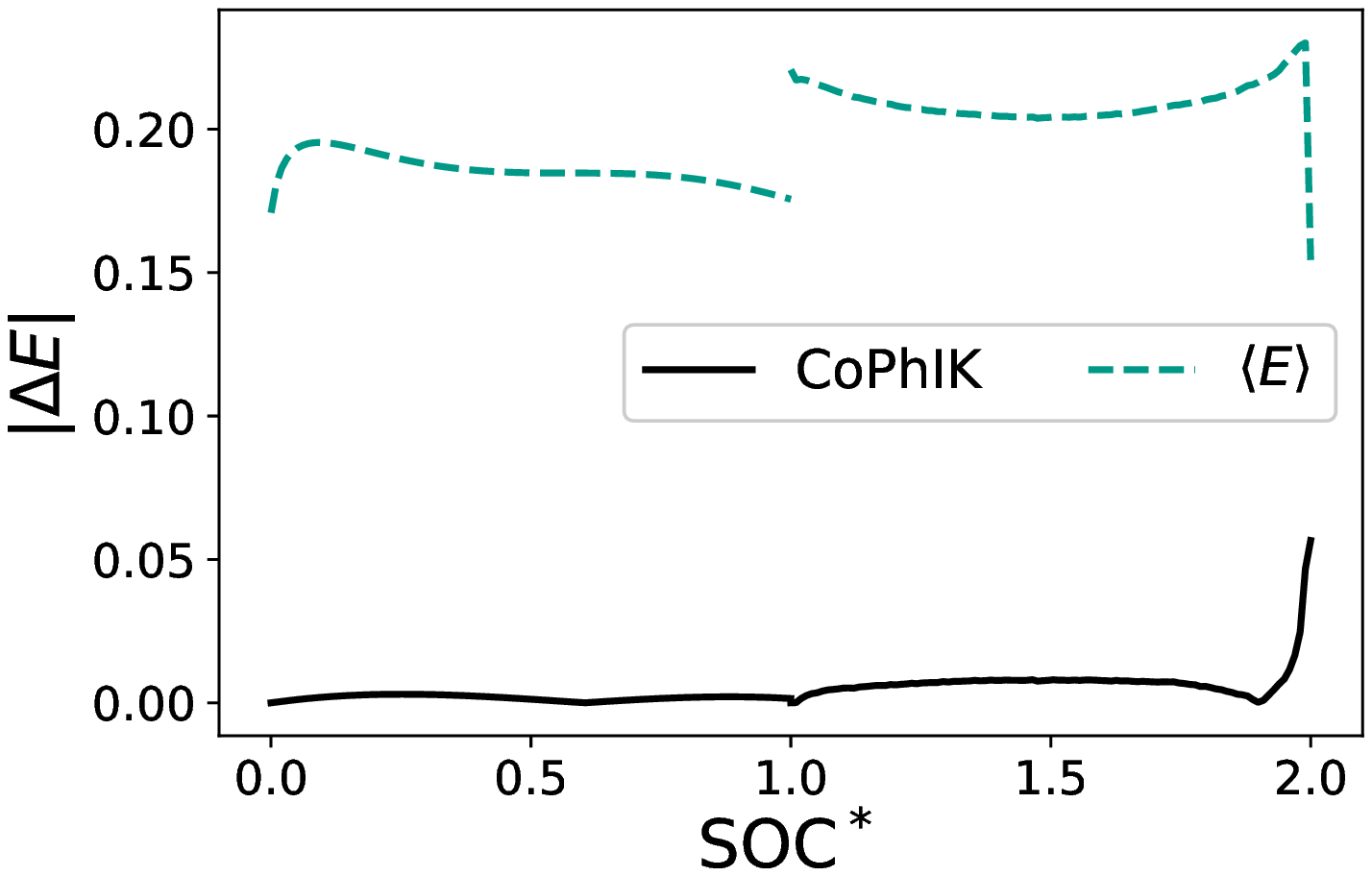}
\includegraphics[width=0.3\linewidth]{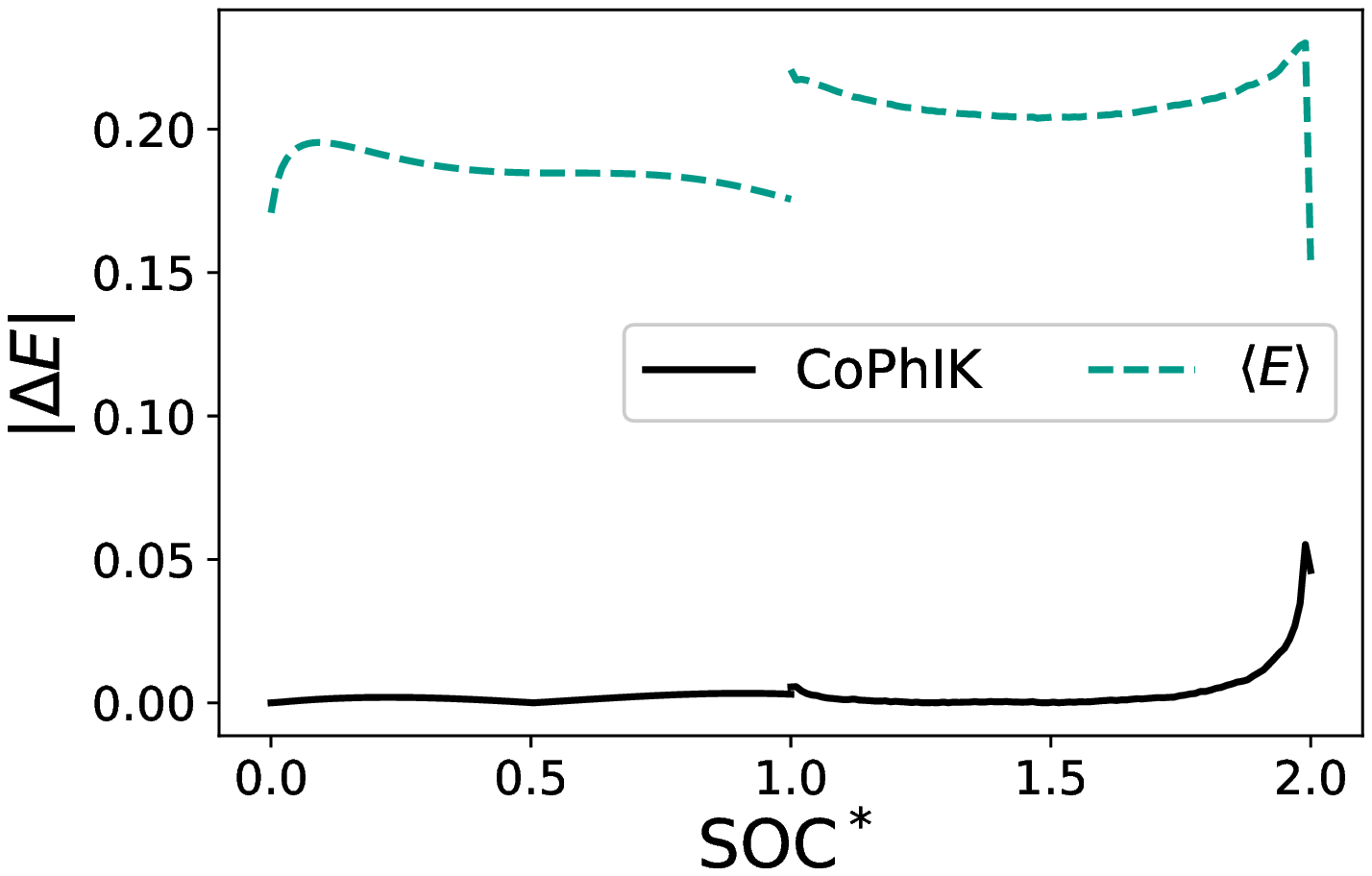}
\caption{Example of CoPhIK results and relative errors for Experiment 2 with $1$, $2$, and $3$ data points included. The top row shows the resulting charge-discharge curve, the bottom row shows the relative errors. The low fidelity dataset is calculated using the values from \cite{Shah2011}. We plot the mean of the low-fidelity dataset ($\langle E \rangle$), the experimental data ($E_{exp}$), and the variance of the CoPhIK results ($E_{var}$).} \label{fig:CoPhIK_Shah}
\end{figure}

In this section we conduct CoPhIK simulations as described above, except the input parameters to the Gaussian random variables in Eqs. (\ref{eq:GRV1})--(\ref{eq:GRV4}) are taken from literature \cite{Shah2011} instead of \cite{Chen2020}. The rest of the data remains the same. Fig. \ref{fig:0d} demonstrates  that $E$ computed from the 0d model with these parameters overestimates the experimental values during charging and underestimates the discharging section by approximately 10$\%$. As a consequence, the mean of the low-fidelity dataset also over- and underestimates the experimental data. However, the results with CoPhIK can compensate for the discrepancies between the experimental and low-fidelity data. In Fig. \ref{fig:CoPhIK_Shah}, we show the 0d and CoPhIK predictions of $E$ for the same experimental data set as in Fig. \ref{fig:CoPhIK1}. The mean of the low-fidelity dataset, $\langle E \rangle$, does not agree well with the experimental data. However, the CoPhIK prediction agrees very well. In some cases, the uncertainty is higher than in Fig. \ref{fig:CoPhIK1}, but the absolute error is of the same order of magnitude with both sets of input parameters to the 0d model. The average $\ell_2$ errors are $4.09 \times 10^{-4}$, $1.64 \times 10^{-4}$, and $9.44 \times 10^{-5}$ with one, two, and three data points, respectively, and the average $\ell_\infty$ errors are $8.85 \times 10^{-2}$, $9.66 \times 10^{-2}$, and $9.99 \times 10^{-2}$ with one, two, and three data points. Comparing with Table \ref{tab:errors}, there is no significant difference between using the input parameters in this work and the input parameters from \cite{Shah2011} for the $\ell_2$ error, and results are similar for the $\ell_\infty$ error. Very similar results are seen when using the parameters from \cite{Eapen2019}. 

These results show that the CoPhIK framework removes the potential bias because of the difference in the prior and actual statistics of the parameters in the 0d model, i.e., the proposed CoPhIK framework is non-sensitive to the choice of the prior statistics of the parameters in the 0d model.

\section{Conclusions}
We have shown that the CoPhIK method can be used to accurately model  RFB systems given a small number of experimental observations. Specifically, we demonstrated that the CoPhIK method accurately predicts the charge-discharge curves of a RFB using only initial conditions from the Supplementary Information and charge-discharge data collected in sixteen experiments with similar but not identical batteries and different initial conditions. We note that the resulting predictions are not sensitive to the prior distribution of the parameters in the 0d model, allowing for accurate predictions without the need to fit the input parameters. As there is disagreement in the literature over what these parameters should be, with suggested values ranging over several orders of magnitude, our model allows for flexibility without the need for expensive and time-consuming calibration procedures, which may require more data than is available. Additionally, the CoPhIK model is able to capture key features of the charge-discharge curve, including the tails that can occur at the beginning and end of each cycle. The 0d model, which does not incorporate data, cannot predict these features. The CoPhIK model is inexpensive to run, with a typical run taking less than a minute. This allows for accurate, rapid prediction of the RFB system's performance. 

In future work we look to use the CoPhIK model to conduct active learning to reduce uncertainty in our model predictions while running minimal expensive experiments. By targeting experiments in areas where the uncertainty is highest, the uncertainty of new predictions can be limited while minimizing the experimental cost and workload. For example, in Fig. \ref{fig:scaling_param}, the uncertainty is highest for small and large initial voltage values. By conducting targeted experiments at higher and lower initial voltages, the uncertainty of the scaling parameter would be minimized. 

\section{Acknowledgements}
The authors thank J. Bao and R. Tipireddy for helpful discussions regarding the 0d modeling and W. Wang and L. Yan for help with interpreting the experimental design and data.
This research was supported by the Energy Storage Materials Initiative (ESMI), under the Laboratory Directed Research and Development (LDRD) Program at Pacific Northwest National Laboratory (PNNL).  PNNL is a multi-program national laboratory operated for the U.S. Department of Energy (DOE) by Battelle Memorial Institute under Contract No. DE-AC05-76RL01830.

\bibliography{ref_battery,Amanda-bib}
\newpage

\section*{Methods}
\subsection{Gaussian process regression}

In GPR we wish to find the cell voltage $E \in \bbR$ at input variables $\mathbf{x}\in \bbD$, $E(\mathbf{x})$,  as a Gaussian process: 

\begin{equation}
    \hat{E}(\mathbf{x}) \sim \mathcal{GP}\left(\mu(\bx), k(\bx, \bx')\right)
\end{equation}
where $\mu(\cdot) : \bbD \rightarrow \bbR$ and and $k(\cdot, \cdot) : \bbD \times \bbD \rightarrow \bbR$ are the mean and covariance function, given by

\begin{align}
    \mu(\bx) &= \mathbb{E}\{\hat{E}(\bx)\} \\
    k(\bx, \bx') &= Cov\{\hat{E}(\bx), \hat{E}(\bx')\} = \mathbb{E}\left\{[\hat{E}(\bx)-\mu(\bx)][\hat{E}(\bx')-\mu(\bx')]\right\},
\end{align}
where the operator $\mathbb{E}\{ \cdot \}$ is the ensemble mean. 
We denote the covariance matrix to be the $N \times N$ matrix given by

\begin{equation}
    \bC = \begin{bmatrix}
    k(\mathbf{x}^{(1)}, \mathbf{x}^{(1)}) & \ldots & k(\mathbf{x}^{(1)}, \mathbf{x}^{(N)}) \\
    \vdots & \ddots & \vdots\\
    k(\mathbf{x}^{(N)}, \mathbf{x}^{(1)}) & \ldots & k(\mathbf{x}^{(N)}, \mathbf{x}^{(N)}) 
    \end{bmatrix}
\end{equation}
and define the covariance vector as 

\begin{equation}
    \bc(\bx^*) = \left( k(\mathbf{x}^{*},\mathbf{x}^{(1)}), \ldots , k(\mathbf{x}^{*}, \mathbf{x}^{(N)})\right)^T.
\end{equation}

The GPR prediction of $E$ at $\bx^*$ is then given by the posterior distribution 
$\hat{E}_p(\bx^*) \sim \mathcal{N} (\mu_p (\bx^*),\sigma^2_p (\bx^*))$, where the posterior mean and variance are given by

\begin{equation}\label{mean}
    \mu_p(\bx^*) = \mu(\bxs) + \bc(\bxs)^T\bC^{-1}(\mathbf{E} - \mathbf{\boldsymbol{\mu}}) 
\end{equation}
\begin{equation}\label{variance}
   \sigma^2_p(\bxs) = \sigma^2(\bxs) - \bc(\bxs)^T\bC^{-1}\bc(\bxs),
\end{equation}
where $\boldsymbol{\mu} = (\mu(\mathbf{x}^{(1)}), \mu(\mathbf{x}^{(2)}), \ldots, \mu(\mathbf{x}^{(N)}))^T$ and $\sigma^2(\bxs) = k(\mathbf{x},\mathbf{x})$. 

 To ensure that $\bC$ is invertible we substitute $\bC = \bC + \epsilon \mathbf{I}$, where $\epsilon \ll 1$. This is equivalent to assuming there is i.i.d. observation noise with variance $\epsilon$ \cite{yang2019}.

\subsection{Monte-Carlo estimates of the mean and covariance of $E$ given by the 0d model}\label{low_fidelity}

The 0d model of an RFB was originally proposed in \cite{Shah2008, Shah2011} and modified by \cite{Eapen2019} to give a fast estimate of redox-flow battery performance. The 0d model gives the cell voltage, $E$, due to an applied current density, $j_{app}$, in the form 

\begin{equation}
E = E_{cell}^{rev} - \eta_{act} - \eta_{ohm},     
\end{equation}
where $E_{cell}^{rev}$ is the reversible Open Current Voltage (OCV), $\eta_{act}$ is the activation overpotential, and $\eta_{ohm}$ is the ohmic losses. A short explanation of the 0d model is given below.

The two half reactions in the RFB are given by

\begin{align}
    V^{3+} + e^- &\rightleftarrows V^{2+} \label{react1}\\
    VO^{2+} + H_2O &\rightleftarrows VO_2^{+} + 2H^+ e^- \label{react2}
\end{align}

From Nernst's equation \cite{Eapen2019}, 

\begin{equation}
    E_{cell}^{rev}(T) = E_2^0(T)-E_1^0(T) + \frac{RT}{F} \ln \left( \frac{C_{V(II)}C_{V(V)}C^2_{H_p^+}C_{H_p^+}}{C_{V(III)}C_{V(IV)}C_{H_2O_p}C_{H_n^+}}\right).
\end{equation} 
where $C_i$ is the molar concentration of species $i$, $E_1^0$ and $E_2^0$ are the formal potentials for the reactions at the negative and positives electrodes, $R$ is the molar gas constant, $T$ is the cell temperature, and $F$ is the Faraday constant. 

The activation overpotential is given by 

\begin{equation}
    \eta_{act} = |\eta_{pos}|+|\eta_{neg}|
\end{equation}

where
\begin{equation}
    \eta_{neg} = -\frac{2RT}{F} a\sinh\left(\frac{j_{app}}{2Fk_1^* \sqrt{C_{V(II)}C_{V(III)}}}\right)
\end{equation}
\begin{equation}
    \eta_{pos} = \frac{2RT}{F} a\sinh\left(\frac{j_{app}}{2Fk_2^* \sqrt{C_{V(IV)}C_{V(V)}}}\right)
\end{equation}
and 

\begin{align}
    k_1^* &= k_{1} \exp \left(- \frac{FE_1^0(T_{ref})}{R}\left[T_{ref}^{-1} - T^{-1}\right]\right)\\
    k_2^* &= k_{2} \exp \left( \frac{FE_2^0(T_{ref})}{R}\left[T_{ref}^{-1} - T^{-1}\right]\right).
\end{align}
The coefficients $k_{1}$ and $k_{2}$ are the reference rate constants for reactions (\ref{react1}) and (\ref{react2}) at $T_{ref} = 293 K$, respectively, and are typically determined from experiments. 

\begin{figure}
\centering
\includegraphics[width=0.4\linewidth]{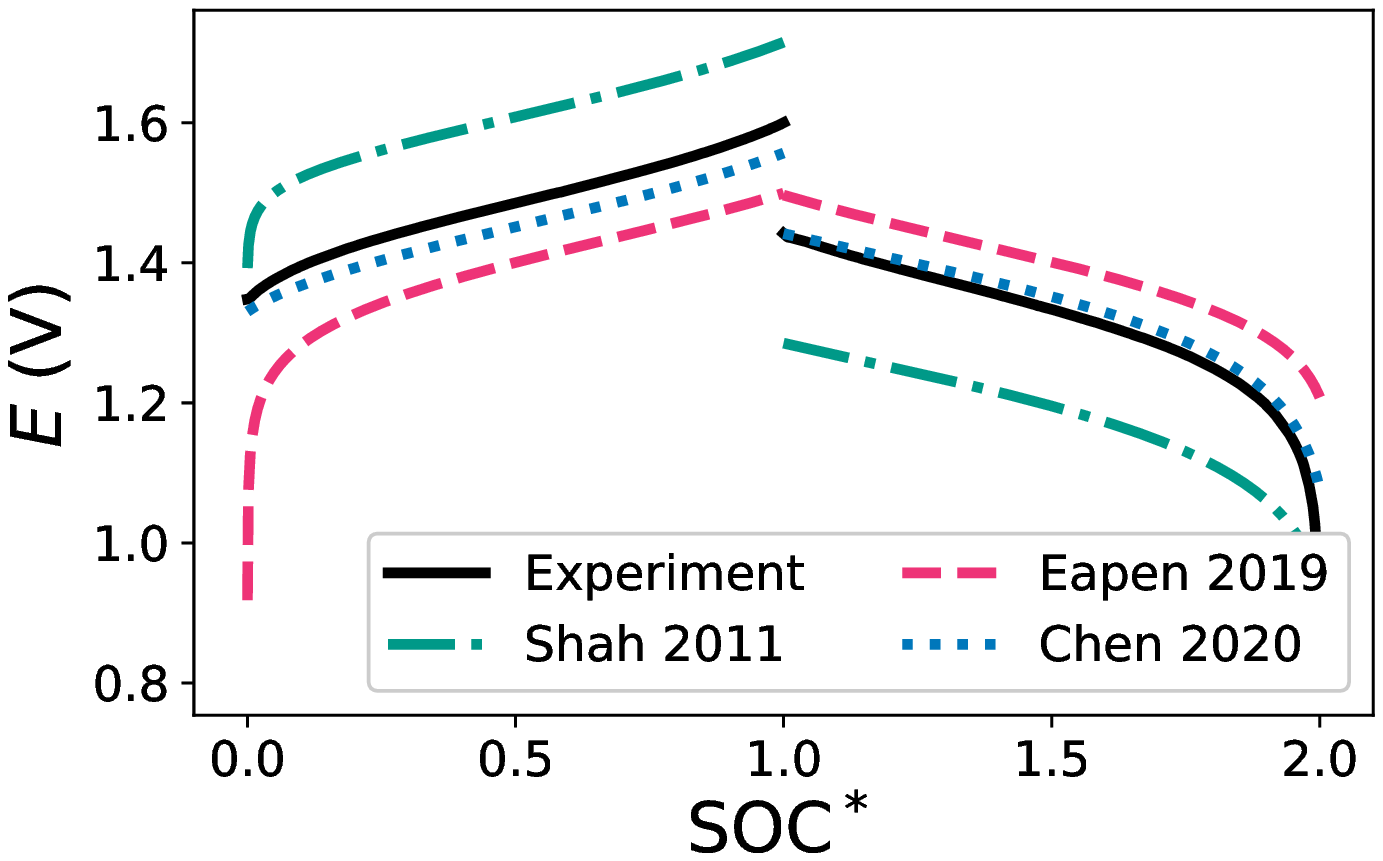}
\includegraphics[width=0.4\linewidth]{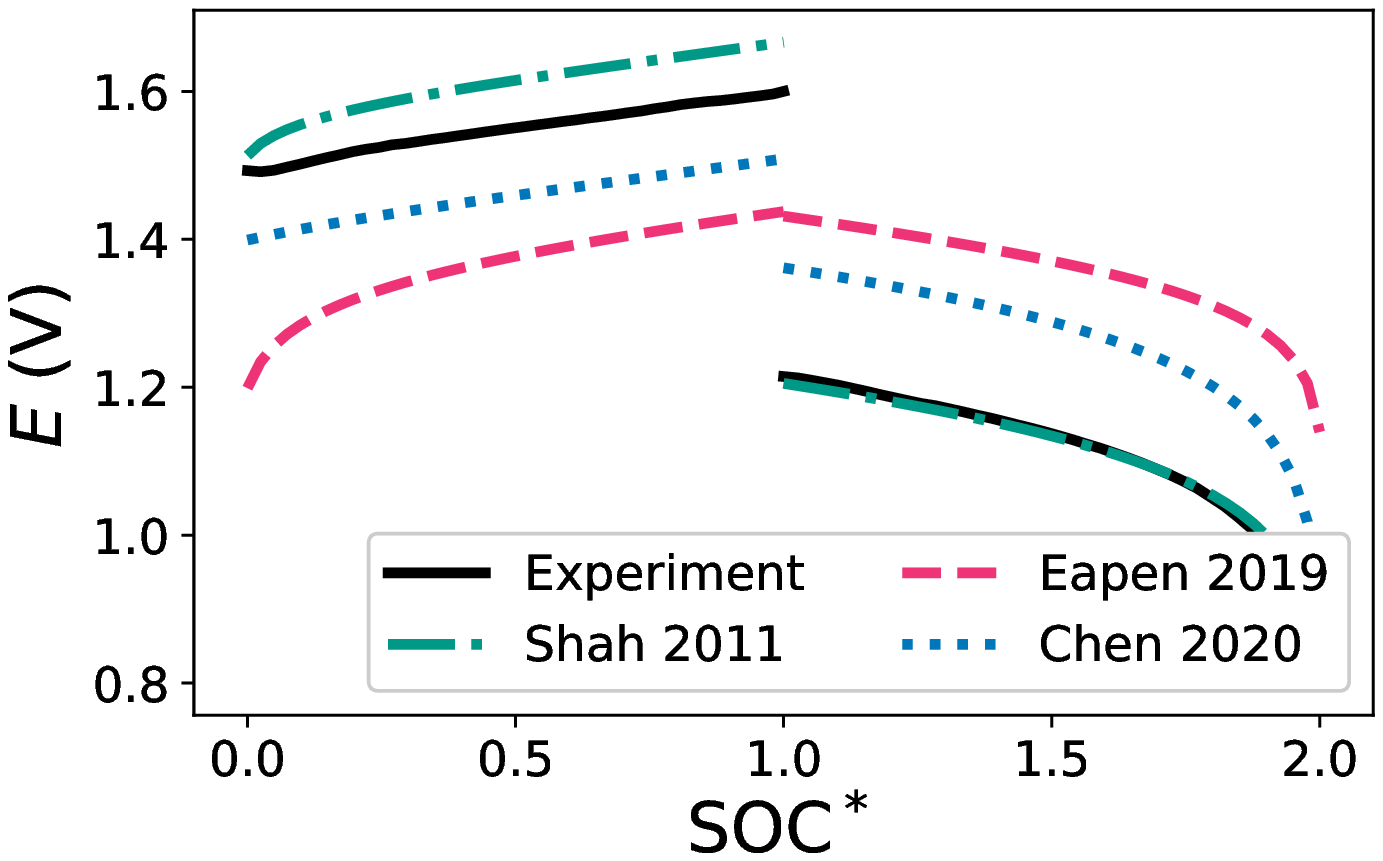}
\caption{Examples of the 0d model with the parameters from \cite{Shah2011, Eapen2019, Chen2020}. }\label{fig:zerod_opt}
\end{figure}
The ohmic losses are given by

\begin{equation}
    \eta_{ohm} = \left(\frac{w_m}{\sigma_m} + \frac{w_e}{\sigma_e \epsilon^{1.5}} + \frac{w_c}{\sigma_c}\right)j_{app}
\end{equation}
with the current collector conductivity $\sigma_c$, the membrane conductivity

\begin{equation}
    \sigma_m = (0.5139 \lambda-0.326)\exp \left(1268\left[303^{-1} - T^{-1}\right]\right)
\end{equation} where $\lambda=22$ is the membrane water content for a fully saturated membrane, 
and the electrolyte ionic conductivity $\sigma_e$ to be determined from experimental results. The widths of the membrane, electrolyte, and current collector are denoted by $w_m$, $w_e$, and $w_c$, respectively. The specific area is given by $A_s = SV_e$, where $S$ is the specific surface area that is determined from experiments and $V_e$ is the electrode volume. Mass balance for each species is used to determine $C_i$ for $i = V(II),$ $V(III),$ $V(IV),$ $V(V),$ $H^+,$ and $H_2O$. For details see \cite{Shah2008, Shah2011}. 

While most of the parameters in the 0d model  can be directly measured or estimated based on the battery design and material properties, the parameters $k_{1}, k_2, \sigma_e$, and $S$ denoting the reference rates for reactions (\ref{react1}) and (\ref{react2}), the electrode conductivity, and the specific surface area, must be calibrated \cite{Shah2011, Eapen2019,Chen2020}. Table \ref{tab:zerod_params} presents values of these parameters from the calibration studies in \cite{Shah2011, Eapen2019,Chen2020}. The fixed parameters are given in Table \ref{tab:zerod_fixed_params}. 

\begin{figure}
\centering
\includegraphics[width=0.4\linewidth]{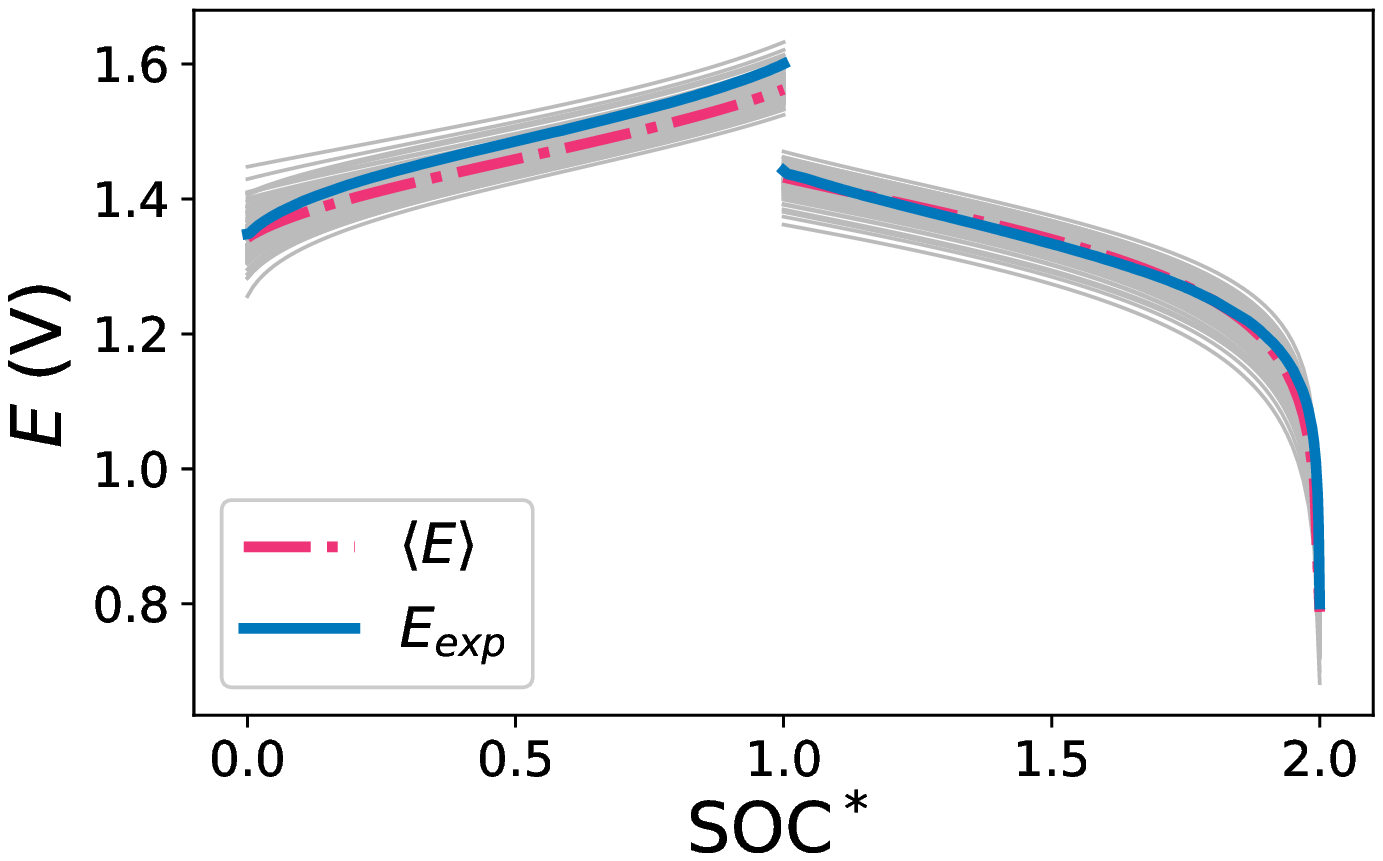}
\includegraphics[width=0.4\linewidth]{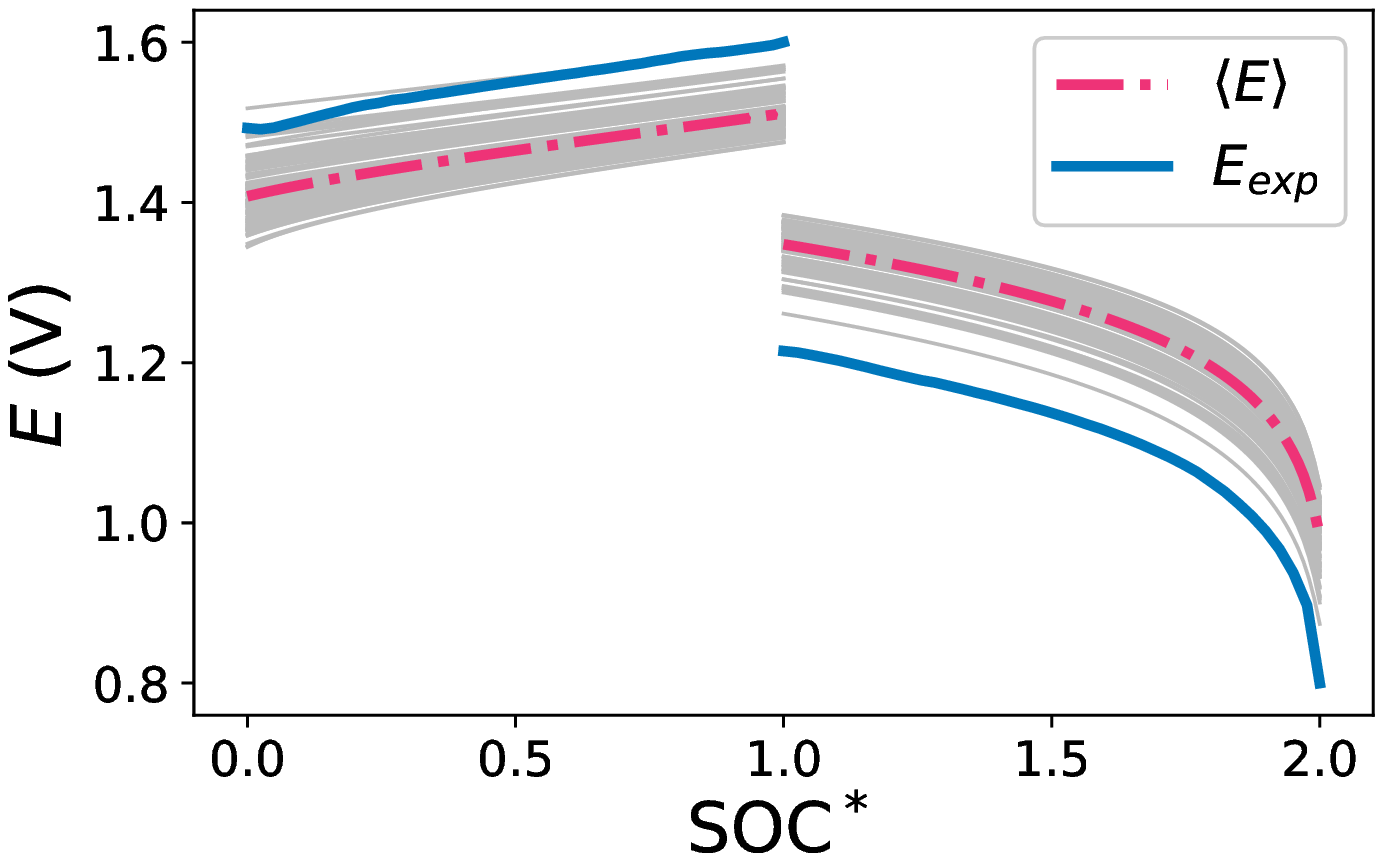}
\caption{Examples of the 0d model generated low-fidelity samples. The experimental results ($E_{exp}$) and the mean of the generated samples ($\langle E \rangle$) are shown. 200 sample profiles are shown for demonstration purposes.} \label{fig:0d}
\end{figure}

\begin{table}
\centering
\begin{tabular}{ c c c c}  \hline
 & \cite{Shah2011} & \cite{Eapen2019} & \cite{Chen2020} \\ \hline
 $k_1$ ($m/s$) & $3.56\times10^{-6}$ & $1.28\times 10^{-5}$ & $5.0 \times10^{-8}$ \\
 $k_2$ ($m/s$) & $3\times10^{-9}$ & $3\times 10^{-5}$ & $1 \times10^{-7}$ \\
 $\sigma_e$ ($s/m$) & 100 & 1000 & 500 \\
 $S$ ($m^2$/$m^3$) & $1.62\times10^4$ & $1.62\times10^4$ & $3.48\times10^4$ \\ \hline
\end{tabular}
\caption{Comparison of the 0d model parameters from \cite{Shah2011}, \cite{Eapen2019}, and \cite{Chen2020}.} \label{tab:zerod_params}
\end{table}
A comparison of the 0d model predictions with these three sets of parameters for two experiments is given in Fig. \ref{fig:zerod_opt}. For all considered experiments, the  parameters in \cite{Eapen2019} underestimate  $E$, while the parameters in \cite{Shah2011} overestimate $E$. The parameters from \cite{Chen2020} were obtained from an experiment in a cell similar to the one used in the experiments considered in this work, and so they provide the best agreement in Fig. \ref{fig:zerod_opt} and serve as the starting point for the MC simulations.

To model the $i$th experiment ($i=1,...,16$), in the MC simulation method, we solve the 0d equations $N_{MC}=1000$ times with the input parameters $$\bx = [t, j, V_r, w_m, C_{V(III)}^0, C_{V(IV)}^0, C_{H^+, pos}^0, C_{H_20, pos}^0, C_{H^+, neg}^0, C_{H_20, neg}^0]$$ taken from Table \ref{tab:HF_params} for experiment $i$. The parameters $\mathbf{v} = [k_1, k_2, \sigma_e, S]^T$ are taken as Gaussian random variables with the mean values $\langle v_i \rangle$ given by \cite{Chen2020} and standard deviation $\sigma_{v_i} = 0.25\langle v_i \rangle$:

\begin{align}
    k_1^m &= \xi^m(k_1, 0.25k_1) \label{eq:GRV1} \\
    k_2^m &= \xi^m(k_2, 0.25k_2)\label{eq:GRV2} \\
    \sigma_e^m &= \xi^m(\sigma_e, 0.25\sigma_e)\label{eq:GRV3} \\
    S^m &= \xi^m(S, 0.25S) \label{eq:GRV4}
\end{align}
where $\xi^m(\mu, \sigma)$ denotes a Gaussian random variable with mean $\mu$ and standard deviation $\sigma.$
Fig. \ref{fig:0d} shows 50 realizations of $E(t)$ computed from the 0d model for one of the experiments as well as $E(t)$ observed in the experiment and computed from the 0d model with $\mathbf{v} = \langle \mathbf{v} \rangle$. 
Because the 0d model involves logarithm functions, for some values of $\mathbf{v}$ the output may be undefined. Therefore, we only consider MC realizations where the value is defined for all time. This figure shows that the mean cell voltage $\mu_L$ obtained from the MC method cannot accurately predict the tails of the charge and discharge curves. Similar results are obtained for the other experiments.   

\begin{table}[h]
\centering
\begin{tabular}{ l l }  \hline
Parameters & Values\\
\hline
Electrode porosity & $\epsilon = 0.67$ \\
Membrane water content & $\lambda = 22$ \\
Current collector conductivity & $\sigma_c = 91000$ s m$^{-1}$ \\
Reference potential, reaction \ref{react1} & $E^0_1 = -0.255 $ V \\
Reference potential, reaction \ref{react2} & $E^0_2 = 1.0 $ V \\
Faraday constant & $F = 96485.3329$ sA/mol \\
Gas constant & $R = 8.314462618$	J/K/mol \\
Reference temperature & $T_{ref} = 293$ K \\
 \hline
\end{tabular}
\caption{Parameters used in the 0d model.} \label{tab:zerod_fixed_params}
\end{table}

\pagebreak
\newpage

\section*{Supplementary information}
\subsection*{Detailed description of high-fidelity data}

  The experiments use Nafion 115 and 212 membranes.  The positive and negative electrodes are made of 3 mm thick graphite felt with active areas of 10 cm$^2$. The electrolytes are prepared by dissolving 1.5 M $VOSO_4$ (Aldrich, 99\%) in 3.5 M $H_2SO_4$ solution (Aldrich, 96–98\%). The electrolytes are placed in two glass reservoirs and circulated at a flow rate of 20 mL/min by a peristaltic pump. The flow cell electrochemical performance are measured and recorded using a potentiostat/galvanostat (Arbin Instrument, USA) with the fixed voltage range of 0.8 to 1.6 V. Table \ref{tab:HF_params} lists parameters that were varied between the experiments while the parameters that were kept constant in all experiments are given in Table \ref{tab:HF_params_same}. The current density in each experiment is kept constant, although it varied between experiences.

\begin{table}[ht]
\centering
\begin{tabular}{ l l}  \hline
Parameters & Values\\
\hline
$v$ & 4.17 m/s \\
$C_{V(II)}(0)$ & 0.0 mol/m$^3$\\
$C_{V(V)}(0)$ & 0.0 mol/m$^3$\\
Temperature & 298.15 K \\
Electrode height & 0.05 m \\
Electrode breadth & 0.025 m \\
Electrode width & 0.004 m \\
Collector width & 0.015 m
 \\ \hline
\end{tabular}
\caption{Initial parameters for the high fidelity data set that are fixed in this paper. } \label{tab:HF_params_same}
\end{table}

\newpage
\begin{turnpage}

\begin{table}
\centering
\hspace*{-1cm}
\begin{tabular}{ c c c c c c c c c c l}  \hline
Exp. & $j$  & $V_r$  & $C_{V(III)}^0$ & $C_{V(IV)}^0$  & $C_{H^+, pos}^0$ & $C_{H_20, pos}^0$  & $C_{H^+, neg}^0$  & $C_{H_20, neg}^0$ & Membrane  & Notes \\
  &  (A) &  (m$^3$) &  (mol/m$^3$) &  (mol/m$^3$)  & (mol/m$^3$) &  (mol/m$^3$) & (mol/m$^3$) &  (mol/m$^3$)  & & \\
\hline
1 & 0.5 & $20 \times 10^{-6}$ &  $1.5 \times 10^3$ & $1.5 \times 10^3$  & $3.85 \times 10^3$ &  $4.46 \times 10^4$ & $3.025 \times 10^3$ &  $4.61 \times 10^4$ &  115 & \\ 
2 & 0.75 & $80 \times 10^{-6}$ &  $1.5 \times 10^3$ & $1.5 \times 10^3$ & $3.85 \times 10^3$ &  $4.46 \times 10^4$ & $3.025 \times 10^3$ &  $4.61 \times 10^4$   &  115 & \\ 
3 & 0.75 & $80 \times 10^{-6}$ &   $1.5 \times 10^3$ & $1.5 \times 10^3$  &  $3.85 \times 10^3$ &  $4.46 \times 10^4$ & $3.025 \times 10^3$ &  $4.61\times 10^4$   &  115 & \\ 
4 & 0.4 & $30 \times 10^{-6}$ &  $1.5 \times 10^3$ & $1.5 \times 10^3$  &  $3.85 \times 10^3$ &  $4.46\times 10^4$ & $3.025 \times 10^3$ &  $4.61 \times 10^4$  &  115 & Sulfuric acid electrolyte\\ 
5 & 0.5 & $50 \times 10^{-6}$ &  $2.0 \times 10^3$ & $2.0 \times 10^3$  &  $5.0 \times 10^3$ &  $4.753 \times 10^4$ & $3.0 \times 10^3$ &  $4.953 \times 10^4$ &  115 & Membrane soaked in $H_2O_2$ for 1 hour  pretreatment \\ 
6 & 0.5 &$50 \times 10^{-6}$ &  $2.0 \times 10^3$ & $2.0 \times 10^3$  &  $5.0 \times 10^3$ &  $4.753 \times 10^4$ & $3.0 \times 10^3$ &  $4.953 \times 10^4$ &  212 & Membrane  soaked in $H_2O_2$ and $O_2$ plasma pretreatment  \\ 
7 & 0.69 &$30 \times 10^{-6}$ &  $2.0 \times 10^3$ & $2.0 \times 10^3$  &  $5.0 \times 10^3$ &  $4.753 \times 10^4$ & $3.0 \times 10^3$ &  $4.953 \times 10^4$ &  115&  \\ 
8 & 0.75 & $45 \times 10^{-6}$ &  $2.0 \times 10^3$ & $2.0 \times 10^3$  &  $5.0 \times 10^3$ &  $4.753 \times 10^4$ & $3.0 \times 10^3$ &  $4.953 \times 10^4$ &  115& \\ 
9 & 0.69 &$45 \times 10^{-6}$ &  $2.0 \times 10^3$ & $2.0 \times 10^3$  &  $5.0 \times 10^3$ &  $4.753 \times 10^4$ & $3.0 \times 10^3$ &  $4.953 \times 10^4$ &  115 & \\ 
10 & 0.75 & $45 \times 10^{-6}$ &  $2.0 \times 10^3$ & $2.0 \times 10^3$  &  $5.0 \times 10^3$ &  $4.753 \times 10^4$ & $3.0 \times 10^3$ &  $4.953 \times 10^4$ &  115 & \\ 
11 & 0.8 & $45 \times 10^{-6}$ &  $2.0 \times 10^3$ & $2.0 \times 10^3$  &  $5.0 \times 10^3$ &  $4.753 \times 10^4$ & $3.0 \times 10^3$ &  $4.953 \times 10^4$ &  212 & Varying temperature \\ 
12 & 0.4 & $50 \times 10^{-6}$ &  $2.0 \times 10^3$ & $2.0 \times 10^3$  &  $5.0 \times 10^3$ &  $4.753 \times 10^4$ & $3.0 \times 10^3$ &  $4.953 \times 10^4$ & 212 & Sulfuric-hydrochloric acid electrolyte\\ 
13 & 0.4 & $25 \times 10^{-6}$ &  $2.0 \times 10^3$ &  $2.0 \times 10^3$  & $5.0 \times 10^3$ &  $4.753 \times 10^4$ & $3.0 \times 10^3$ &  $4.953 \times 10^4$ &  212 & 10\% of $NH_4Cl$\\ 
14 & 0.5 & $40 \times 10^{-6}$ &  $2.0 \times 10^3$ &  $2.0 \times 10^3$  & $5.0 \times 10^3$ &  $4.753 \times 10^4$ & $3.0 \times 10^3$ &  $4.953 \times 10^4$ &  212& Acetic acid additive\\ 
15 & 0.5 & $40 \times 10^{-6}$ &  $2.0 \times 10^3$ &  $2.0 \times 10^3$  & $5.0 \times 10^3$ &  $4.753 \times 10^4$ & $3.0 \times 10^3$ &  $4.953 \times 10^4$ &  212 & \\ 
16 & 0.5 & $40 \times 10^{-6}$ &  $2.0 \times 10^3$ &  $2.0 \times 10^3$  & $5.0 \times 10^3$ &  $4.753 \times 10^4$ & $3.0 \times 10^3$ &  $4.953 \times 10^4$ &  212& Different electrode from Toyobo \\ 
 \hline
\end{tabular}
\caption{Initial parameters for the high fidelity data set considered in this paper. We consider the current $j$, the reservoir volume $V_r$, the membrane, and the initial concentrations of $V(III)$, $V(IV)$, $H^+$, and $H_2O$, denoted by $C_i^0$ for $i = V(III)$, $V(IV)$, $H^+$, and $H_2O$. For $H^+$, and $H_2O$ the $pos$ and $neg$ subscripts denote the initial concentrations in the positive and negative reservoirs, respectively. The membranes used are Nafion 115 and Nafion 212, and the membrane width is $w_m = 1.27 \times 10^{-4}$ m for Nafion 115 and $w_m = 5.08 \times 10^{-5}$ m for Nafion 212.} \label{tab:HF_params}
\end{table}

\end{turnpage}
\end{document}